\documentclass[twocolumn]{aastex631}

\newcommand{\btheta}{\boldsymbol {\theta}}
\newcommand{\bphi}{\boldsymbol {\phi}}
\newcommand{\bx}{\mathbf{x}}

\newcommand{\err}[2][]{#2_{\text{err}#1}}
\newcommand{\bxerr}{\err{\bx}}
\newcommand{\bxerri}{\err[,i]{\bx}}
\newcommand{\yerri}{\err[,i]{y}}

\newcommand{\obs}[2][]{#2_{\text{obs}#1}}

\newcommand{\bxobsi}{\obs[,i]{\bx}}
\newcommand{\yobsi}{\obs[,i]{y}}

\newcommand{\linmix}{{\tt LINMIX}}
\newcommand{\ltsfit}{{\tt LtsFit}}
\newcommand{\leopy}{{\tt Leopy}}
\newcommand{\mlmethod}{{ML based method}}
\newcommand{\ksmethod}{{KS-test based method}}
\newcommand{\roxy}{{\tt ROXY}}
\newcommand{\tcpu}{{\tt t-cup}}
\newcommand{\raddest}{{\tt raddest}}

\newcommand{\python}{{\tt Python}}

\newcommand{\pzflow}{{\tt PZFlow}}
\newcommand{\R}{{\tt R}}

\newcommand{\yaImpute}{{\tt yaImpute}}
\newcommand{\energy}{{\tt energy}}
\newcommand{\cramer}{{\tt cramer}}

\newcommand{\tdks}{2D KS test}

\usepackage{amsmath}

\shorttitle{Regression for Astronomical Data}
\shortauthors{Jing \& Li}
\graphicspath{{./}{figures/}}

\begin{document}

\title{Regression for Astronomical Data with Realistic Distributions, Errors and Non-linearity}

\correspondingauthor{Tao Jing}
\email{jingt20@mails.tsinghua.edu.cn}
\correspondingauthor{Cheng Li}
\email{cli2015@tsinghua.edu.cn}

\author[0009-0004-6271-4321]{Tao Jing}
\affiliation{Department of Astronomy, Tsinghua University, Beijing 100084, China}

\author[0000-0002-8711-8970]{Cheng Li}
\affiliation{Department of Astronomy, Tsinghua University, Beijing 100084, China}



\begin{abstract}
We have developed a new regression technique, the maximum likelihood (ML)-based method and its variant, the \ksmethod, designed to obtain unbiased regression results from typical astronomical data. A normalizing flow model is employed to automatically estimate the unobservable intrinsic distribution of the independent variable as well as the unobservable correlation between uncertainty level and intrinsic value of both independent and dependent variables from the observed data points in a variational inference based empirical Bayes approach. By incorporating these estimated distributions, our method comprehensively accounts for the uncertainties associated with both independent and dependent variables.  Our test on both mock data and real astronomical data from PHANGS-ALMA and PHANGS-JWST demonstrates that, given a sufficiently large sample size ($>$ 1000), both the \mlmethod\ and the \ksmethod\ significantly outperform the existing widely-used methods, particularly in cases of low signal-to-noise ratios. The \ksmethod\ exhibits remarkable robustness against deviations from underlying assumptions, complex intrinsic distributions, varying correlations between uncertainty levels and intrinsic values, inaccuracies in uncertainty estimations, outliers, and saturation effects. For sample sizes between 300 and 1000, the \mlmethod\ yields the best performance. In the low data regime ($<$ 300), the \mlmethod\ maintains comparable performance to other state-of-the-art methods. A GPU-compatible Python implementation of our methods, nicknamed ``\raddest'', will be made publicly available upon acceptance of this paper. 
\end{abstract}



\section{Introduction} \label{sec:intro}

Linear and log-linear regression analyses are extensively applied in astronomical research. As a generalization of the conventional Ordinary Least Squares (OLS) estimator, the Weighted Least Squares (WLS) method incorporates the unequal variance of data points (heteroscedasticity) into the regression. Both OLS and WLS are derived with assumption that the independent variables have no observational uncertainties. However, this assumption is generally invalid for astronomical data. The presence of uncertainties in both dependent and independent variables can lead to a bias towards zero in the estimated slope of the (log-)linear relation \citep[e.g.][]{Fuller2009}. Therefore, Orthogonal Distance Regression (ODR) and similar techniques have been introduced to mitigate the bias \citep[e.g.][]{Isobe1990, Hogg2010, Robotham2015}. While ODR yields satisfactory results when its explicit and implicit assumptions are met, substantial accuracy is not guaranteed in some complex, yet not uncommon cases. In addition, OLS and ODR often produce significantly different results \citep[e.g.][]{Ellison2021}, making the selection of regression techniques a non-trivial issue. 

In the past three decades, considerable efforts have been dedicated to develop regression techniques for astronomical data \citep[e.g.][]{Press1992, Akritas1996, Kelly07, Hogg2010, ltsfit2013, Robotham2015, Feldmann19, Bartlett2023,Martin2024}. Although providing significantly improved regression results in many cases, these methods have various limitations and still require further improvement in certain cases. For instance, when deal with log-linear regression, the widely-applied \linmix\ method \citep{Kelly07} relies on the ``delta method'' to convert uncertainties in variables to uncertainties in their logarithms, an approximation method that is accurate only for data of high signal-to-noise ratios (S/N). In order to address the uncertainties in an independent variable, one needs to model the intrinsic distribution of the variable in the first place. This is done by employing Gaussian Mixture Models (GMMs) in both \linmix\ and \roxy, a recent JAX-based implementation of a similar technique that supports non-linear cases \citep{Bartlett2023}, as well as in \tcpu, another recent work that replaces the Gaussian distribution with Student's t distribution to model conditional intrinsic scatter for enhanced robustness \citep{Martin2024}. While GMMs offer a workable solution, they become inefficient when the intrinsic distribution of the variable is complex. 

Furthermore, GMMs introduce an additional hyperparameter, the number of Gaussian components, which needs to be determined. \cite{Bartlett2023} show analytically that a single Gaussian is sufficient, regardless of the actual intrinsic distribution of the independent variable, given an infinite sample size and constant uncertainties. However, these assumptions are not always valid for astronomical datasets, which often exhibit heteroscedasticity and more complex noise behaviors than assumed. The Gaussian distribution decays rapidly away from its mean, leading to a significant gradient vanishing problem which makes optimization difficult for gradient-based algorithms. Consequently, these methods rely on hierarchical Bayesian approaches, using Markov Chain Monte Carlo (MCMC, in \linmix) or Hamiltonian Monte Carlo (HMC, in \roxy) to simultaneously sample the parameters of the GMM and the regression parameters, thereby providing a complete posterior distribution. While offering a comprehensive description of the results, this approach significantly increases computational costs. An alternative method based on extreme deconvolution \citep{Bovy2011} is implemented in \tcpu, which still utilizes GMMs as the underlying model but optimizes it from observed data points using the Expectation–maximization algorithm. Additionally, these methods neglect the potential correlation between the uncertainty level of a variable and its intrinsic value (the value with no observational error), which could arise due to the Poisson nature of photon counting and observational strategies that aim for fixed S/N. 

Another widely-applied method, \leopy\ \citep{Feldmann19} addresses arbitrary correlation forms, dependence between data points, and the presence of censored or missing data. In this method, however, the intrinsic distribution of the independent variable is not modelled natively, but is required to be provided as user input. This presents a challenge in real-world applications where the intrinsic distribution is often unknown.

To address these limitations, we propose a new regression technique that combines the approaches of \cite{Kelly07}, \cite{Feldmann19} and \cite{Bartlett2023}, the advanced machine learning technique of Normalizing Flows \citep[NFs; e.g.,][]{Dinh2014, Rezende2015}, and variational inference-based empirical Bayes analysis. This new technique employs NF to model both the intrinsic distribution of the independent variable and the correlation between uncertainty level and intrinsic value. These distributions are estimated directly from the observed data points. Incorporating these estimated distributions allows for a comprehensive consideration of uncertainties in both independent and dependent variables. Our method can be applied to problems exhibiting intrinsic scatter and non-linear relationships. Furthermore, to enhance robustness, we introduce a variant version that replaces the likelihood (or posterior) with a distribution distance as the optimization objective. 

We evaluate our proposed methods and compare them with other commonly used techniques in the context of log-linear regression (distinct from linear correlation, as we will demonstrate). Our analysis utilizes both simulated data exhibiting log-linear relationships and real astronomical data from PHANGS-ALMA \citep{Leroy2021_PHANGS_ALMA_pipeline, Leroy2021_PHANGS_ALMA_survey} and PHANGS-JWST \citep{Lee2023_PHANGS_JWST, Williams2024_PHANGS_JWST}, specifically focusing on the log-linear correlation between the CO(2-1) emission line flux and the JWST mid-infrared fluxes observed with the F770W, F1000W, F1130W, and F2100W filters. Tests on both simulated and real data demonstrate that, given a sufficient sample size ($>$ 1000), our proposed method, particularly the distribution distance based variant, exhibits superior overall performance and enhanced robustness. For sample sizes between 300 and 1000, the proposed likelihood-based method still outperforms other methods. In the low-data regime ($<$ 300), the proposed likelihood-based method achieves comparable performance to other state-of-the-art algorithms, while its high efficiency, owing to GPU acceleration, still makes it a competitive option. While the experiments on simulated and real data focus on one-dimensional regression problems, our method is derived within the framework of multidimensional regression (as detailed in \autoref{sec:intro_method}). Given NF's effectiveness in modeling high-dimensional distributions, it would be straightforward to generalize our method for regression of higher dimensions.

This paper is organized as follows. In \autoref{sec:methods} we describe both the proposed method and other methods from previous studies. \autoref{sec:experiment} showcases the test in log-linear case on mock data. Next, \autoref{sec:real_data} presents the application of the method to real PHANGS-ALMA and PHANGS-JWST data.  Finally, we discuss and summarize our results in \autoref{sec:dicussion} and \autoref{sec:summary}. 

\section{Regression methods}\label{sec:methods}

In this section, we firstly derive the statistical model for the new regression method proposed in this work. We then describe the methods widely-used in previous studies, and discuss on the connection and differences of our methods with respect to the previous methods. 

\subsection{The new regression method} \label{sec:intro_method}


We derive the statistical model for our regression method in the framework of maximum a posteriori probability (MAP) estimation. This approach can be viewed as a regularization of the maximum likelihood (ML) estimation, incorporating prior information. Our goal is to infer the posterior probability of model parameters given the observed data. According to Bayes theorem, this posterior probability is given by 
\begin{equation}
    P(\btheta| D) = \frac{P(D | \btheta) P (\btheta)}{P(D)},
\end{equation}
where $D$ represents the observed data and $\btheta$ denotes the model parameters. The observed data is a set of data points, i.e. $D=\{D_i\}$, $ i=1,2,...m$, assumed to be independent of each other\footnote{This assumption is not always held in real data, as discussed in \autoref{sec:dicussion}.}. In this case, the likelihood of the observed data given the model parameters can be rewritten as:
\begin{equation}\label{eq:ind_assumption}
    P(D | \btheta) = P(\{D_i\}|\btheta) = \prod_i^m P(D_i|\btheta).
\end{equation}
Here, each data point $D_i$ consists of a one-dimensional dependent observed variable $\yobsi$, an $n$-dimensional independent observed variable $\bxobsi$, and their corresponding uncertainty levels, $\yerri$ and $\bxerri$. The likelihood of data point $D_i$ given the parameters $\btheta$ is thus derived by
\begin{equation} \label{eq:integral}
    \begin{aligned}
    P(D_i|\btheta) = & P(\yobsi, \yerri, \bxobsi, \bxerri|\btheta) \\
         = & \int P(\yobsi, \bxobsi|\yerri, \bxerri, y, \bx)        \\
         & \times P(\yerri, \bxerri|y, \bx) \ P(y, \bx|\btheta)\ dy\ d\bx,
    \end{aligned}
\end{equation}
where $P(y,\bx|\btheta)$ is the underlying model for the intrinsic variables $y$ and $\bx$, and $P(\yobsi, \bxobsi|\yerri, \bxerri, y, \bx)$ and $P(\yerri, \bxerri|y, \bx)$ encapsulate the error properties. This paper focuses on the case of Gaussian errors.  Therefore, the integral in \autoref{eq:integral} can be evaluated using Gauss-Hermite quadrature.  It is worth noting that, in general cases, Monte Carlo integration with importance sampling provides a viable alternative. Throughout this work we employ Gauss-Hermite quadrature to handle integrals unless otherwise stated.

The underlying model of the intrinsic variables  $P(y, \bx|\btheta)$ 
 can be decomposed as:
\begin{equation} \label{eq:P_xy}
    P(y, \bx|\btheta) =
    P(y |\bx, \btheta) P(\bx),
\end{equation}
where $P(y |\bx, \btheta)$ describes how $\bx$ determines $y$ given $\btheta$. Apparently, $P(\bx)$, the intrinsic distribution of $\bx$ contributes to the likelihood function, and its contribution cannot be simply ignored unless the uncertainties in the observed data are negligible. Similarly, the likelihood of the errors can also be decomposed if the uncertainty levels of $y$ and $\bx$ are independent of each other \footnote{This is not the same as independence of the noise in $y$ and $\bx$, which is represented by  $P(\yobsi, \bxobsi|\yerri, \bxerri, y, \bx) = P(\yobsi|y, \yerri)P(\bxobsi|\bx, \bxerri)$ and is not required, as discussed in \autoref{sec:dicussion}}:
\begin{equation} \label{eq:P_xerr_x}
    P(\yerri, \bxerri|y, \bx) = P(\yerri|y) P(\bxerri|\bx).
\end{equation}
The correlation between $\bx$ and $\bxerr$ as described by $P(\bxerri|\bx)$ cannot be simply ignored as well.

In most cases, however, both $P(\bx)$ and $P(\bxerri|\bx)$ are unknown. In this work, we model these unknown distributions using the machine learning technique of normalizing flows \citep[NFs;][]{Dinh2014,Rezende2015}. As a generative model, a NF learns a bijective transformation $f_{\phi}: x \rightarrow z$, where $x$ represents the variable whose distribution we aim to model and $z$ is a latent variable following a predefined distribution $P_z$. The probability density function of $x$ can be expressed as:
\begin{equation} 
	P_{\phi}(x) = P_z\left(f_\phi(x)\right)\left|\det\left(J_{f_\phi(x)}\right)\right|,
\end{equation} 
where $J_{f_\phi(x)}$ denotes the Jacobian of $f_{\phi}$. For our case, specifically, the probability density function to be modeled  is 
\begin{equation}
P(\bxerri, \bx; \bphi) = P(\bxerri|\bx; \bphi_1) P(\bx; \bphi_2).
\end{equation}
Therefore, in addition to the parameters $\btheta$ that we aim to estimate, two additional sets of nuisance parameters, $\bphi_1$ and $\bphi_2$, that describe $P(\err{\bx}|\bx; \bphi_1)$ and $P(\bx; \bphi_2)$ are required to calculate \autoref{eq:P_xerr_x} and \autoref{eq:P_xy}. To decouple the estimation of $\bphi$ (representing the combination of $\bphi_1$ and $\bphi_2$) and $\btheta$, we estimate $P(\err{\bx}|\bx; \bphi_1)$ and $P(\bx; \bphi_2)$ by maximizing the following likelihood function, which is independent of $\btheta$:
\begin{equation} \label{eq:NF_like}
    P(\{\bxobsi\}, \{\bxerri\}) = \prod_i P (\bxobsi, \bxerri;\bphi)
\end{equation}
where
\begin{equation} \label{eq:NF_like_int} 
    \begin{aligned}
        P (\bxobsi, \bxerri; \bphi) = 
        \int P(\bxobsi|\bx, \bxerri) P(\bxerri, \bx; \bphi) d\bx
        \\
        =\int P(\bxobsi|\bx, \bxerri) 
        P(\bxerri|\bx; \bphi_1)
        P(\bx; \bphi_2)\ d\bx.
    \end{aligned}
\end{equation}
The integral on the right-hand side of \autoref{eq:NF_like_int} 
is handled using Gauss-Hermite quadrature. This approach can be viewed as a form of variational inference based empirical Bayes analysis, aiming to estimate the unobservable population distribution from the observed dataset. The intrinsic distribution and conditional uncertainty level distribution of the dependent variable $y$, $P(y)$ and $P(\err{y}|y)$, can be estimated using the same method. With a well-trained NF, one can not only compute the likelihood of any given $x$, but also generate random variables sampled from the distribution $P_\phi(x)$, which is $P(\bxerri, \bx; \bphi$) or $P(\yerri, y; \bphi)$ in this work.

When using NFs, one challenge is to determine whether the model is adequately fitted to the data. To address this issue, we employ the fitted NF model to generate synthetic data $(\bx_{\rm gen}, \err[,\rm gen]{\bx})$. We then introduce noise to $\bx_{\rm gen}$ according to the error model (a Gaussian distribution in this work) and corresponding $\err[,\rm gen]{\bx}$, obtaining $\obs[,\rm gen]{\bx}$. We assess the goodness of fit for the NF model by comparing the distributions of the generated dataset $(\obs[,\rm gen]{\bx}, \err[,\rm gen]{\bx})$ and observed dataset $(\obs{\bx}, \err{\bx})$. This comparison is evaluated using the two-dimensional Kolmogorov–Smirnov (2D KS) test \citep{Peacock1983, Fasano1987, Press2002}. With the well-fitted NF model, we calculate the integral in \autoref{eq:integral} and subsequently determine the likelihood function $P(D|\btheta)$ through \autoref{eq:ind_assumption}. 

We note that the 2D KS test lacks a rigorous mathematical foundation, and several alternative methods are available \citep{Barnett1976, Rizzo2019}. These methods, which are more robust for evaluating goodness-of-fit, can be implemented using \R\ packages such as \yaImpute, \energy, and \cramer\ \citep{Franz2006, yaImpute2007, Feigelson2012, Rizzo2022}. However, as discussed in the next paragraph, distribution distance measurements are also applied in iterative optimization. Thus, it is essential to balance accuracy and efficiency. In this context, the 2D KS test, which provides an approximate analytical formula for the p-value (as opposed to alternative methods that rely on computationally expensive permutations or bootstrapping), achieves a reasonable trade-off between these factors. For consistency, we use the 2D KS test to evaluate goodness-of-fit in this study. If higher accuracy is required in practical applications, the 2D KS test can be easily replaced with alternative methods. 

Beyond standard maximum MAP or ML estimation, the generative capabilities of NF models allow for parameter estimation by minimizing the distribution distance. In this approach, we first generate $(\bx_{\rm gen}, \obs[,\rm gen]{\bx}, \err[,\rm gen]{\bx})$ as previously described. We then generate $y_{\rm gen}$ according to $P(y|\bx, \btheta)$, followed by $\err[,\rm gen]{y}$ based on $P(\err{y}|y)$. In the next step, noise is introduced to $y_{\rm gen}$ according to $\err[,\rm gen]{y}$. Finally, we use the $p$-value from the \tdks\ between $(\obs[,\rm gen]{\bx}, \obs[,\rm gen]{y})$ and $(\obs{\bx}, \obs{y})$ as the optimization objective to find the best $\btheta$. In the following sections, we refer to the ML(or MAP with uniform prior)-based estimation as the ``\mlmethod'' and the 2D KS-based estimation as the ``\ksmethod''. The \ksmethod\ demonstrates greater robustness than the standard \mlmethod, as to be shown in \autoref{sec:experiment}.

For the \mlmethod, we employ a gradient descent algorithm to find the ML or MAP solution, with the gradient calculated via automatic differentiation. For the \ksmethod, given that the \tdks\ is not amenable to automatic differentiation, we apply a grid search method to determine the maximum p-value solution. Specifically, we first generate a coarse grid covering the initial-guess interval (prior) of the parameters and calculate the p-value at each node. We then calculate the weighted average and standard deviation, using the p-value as the weight. New grids are generated, covering the 2$\sigma$ range. We repeat this approach until the weighted average converges. Finally, we use this converged weighted average as the maximum p-value solution.

For the \mlmethod, we have two options for generating data points from the posterior distribution. The first is a standard MCMC or HMC sampling approach. The second is to generate samples based on the prior distribution and calculate the likelihood as weights, thereby obtaining a weighted sample that represents the posterior distribution. This method leverages the parallel processing capabilities of GPUs and yields samples with equivalent statistical power to MCMC- or HMC-based samples. Therefore, although MCMC sampling is also implemented, we adopt this weighted strategy in this work. In the case of the \ksmethod, obtaining a rigorous posterior sample is infeasible. Consequently, we generate data points from the prior distribution and utilize the p-value provided by \tdks\ as an approximate weight for each data point.

We should point out that, similar formulas to \autoref{eq:integral} have been derived in previous works \citep[e.g.][]{Kelly07, Feldmann19, Bartlett2023,Martin2024}. Compared to those works, our approach has several key improvements. First, our method is not restricted to linear regression problems, unlike the approach presented in \cite{Kelly07}. Next, we employ an NF framework to estimate $P(\bx)$ directly from the data, unlike the approach of \cite{Feldmann19} which requires the user to provide this information as input. In addition, the NF approach offers greater efficiency and flexibility compared to the GMM used in \cite{Kelly07}, \cite{Bartlett2023} and \cite{Martin2024} to model $P(\bx)$. This advantage is demonstrated in the subsequent analysis and in \cite{Yan2025}. Finally, our method accounts for the correlation between $\bx$ and $\err{\bx}$, a common feature of astronomical data due to the Poisson nature of photon counts and the prevalent use of S/N based observation strategies. These improvements lead to enhanced performance compared to previous works, especially for datasets with limited S/N, as to be demonstrated below in \autoref{sec:experiment}.  

\begin{deluxetable*}{ccccc}
	\tablecolumns{5}
	\tablewidth{0pt}
	\tablecaption{Comparison of Different Regression Methods \label{tab:method_compare}}
	\tablehead{
		\colhead{Method} &
		\colhead{$P(\bx)$} &
		\colhead{$P(\bx_{err}|\bx)$ \tablenotemark{a}} &
		\colhead{$P(y|\bx,\btheta)$ \tablenotemark{b}} &
		\colhead{Optimization Objective}
	}
	\startdata
	\mlmethod\ (This work) & NF & NF & Any(Any) & Likelihood/Posterior \\
	\ksmethod\ (This work) & NF & NF & Any(Any) & $p$-value of \tdks \\
	OLS/WLS & ... & ... & Linear(...) & Likelihood \\
	ODR/wODR & Uniform & ... & Linear(...) & Likelihood \\
	mODR & ... & ... & Linear(...) & Likelihood \\
	\linmix/\roxy/\tcpu & GMM & ... & Linear(G)/Any(G)/Linear(T) & Posterior \\
	\leopy\ & Input (heuristic) & ... & Any(Any) & Likelihood/Posterior \\
	\ltsfit\ & ... & ... & Linear(G) & Likelihood \\
	\enddata
	\tablenotetext{a}{$P(\err{y}|y)$ is modelled by the same method as $P(\bx_{err}|\bx)$.}
	\tablenotetext{b}{The content preceding the parentheses indicates the correlation between  $E(y)$ and $x$, while the content within the parentheses describes the model for conditional intrinsic scatter: ``G'' for Gaussian distribution and ``T'' for Student's t distribution.}
\end{deluxetable*}

\subsection{Previous regression methods}

In the next section we will perform the log-linear regression, a common problem in astronomical research, as a test case to evaluate the \mlmethod\ and the \ksmethod. For comparison, we will  also use the methods described in the following. These are  commonly used by astronomical community. 
\begin{enumerate}
    \item Ordinary Least Squares (OLS) regression: This method assumes the accuracy measurement of independent variable and the homoscedasticity of dependent variable. A best-fitted line is obtained by minimizing the sum of squared residuals.
    
    \item Weighted Least Squares (WLS) regression: This method is similar to OLS, but considers the heteroscedasticity of dependent variable. 
    
    \item Orthogonal Distance Regression (ODR): This method minimizes the perpendicular distance between the data points and the fitted line, effectively accounting for uncertainties in both $x$ and $y$. However, it assumes equal variances and homoscedasticity for $\obs{x}$ and $\obs{y}$, and does not utilize $\err{x}$ and $\err{y}$.
    
    \item Weighted Orthogonal Distance Regression (wODR): This method is similar to ODR but utilizes $\err{x}^2$ and $\err{y}^2$ as the variances of $\obs{x}$ and $\obs{y}$.
    
    \item Median Orthogonal Distance Regression (mODR): This method, inspired by \cite{Leroy2023}, first splits the data into several $x_{obs}$ bins, then calculates the median of $x_{obs}$ and $y_{obs}$ in each bin, and finally applies ODR to these medians.
    
    \item \linmix: This Bayesian linear regression code, developed by \cite{Kelly07}, accounts for uncertainties in both $x_{obs}$ and $y_{obs}$. We employ the \python\ implementation of \linmix\footnote{https://github.com/jmeyers314/linmix}. We note the recent advancements in efficiency and flexibility offered by \roxy\ \citep{Bartlett2023} and robustness offered by \tcpu\ \citep{Martin2024}.  Despite these improvements, we utilize \linmix\ in this experiment due to its widespread use in previous research and its approximate equivalence to \roxy\ and \tcpu\ in normal case. 
    
    \item \ltsfit\ \citep{ltsfit2013}: This method, based on the robust Least Trimmed Squares (LTS) technique, addresses regression problems with intrinsic scatter and potential outliers.  We utilize the official \python\ implementation\footnote{https://pypi.org/project/ltsfit/}.
    
    \item \leopy: This method, developed by \cite{Feldmann19}, handles complex noise behavior in astronomical data and supports arbitrary relationships between dependent and independent variables. The official \python\ implementation\footnote{https://github.com/rfeldmann/leopy} is used in this work.
    
\end{enumerate}

It is noteworthy that the log-linear case differs from the linear case, where both the linear relationship and Gaussian error are in linear space. None of the methods listed above, except for \leopy, \roxy\ and the two methods we propose, directly address this log-linear problem. For the other methods, we use $\log \obs{x}$ and $\log \obs{y}$ in regression and employ the ``delta method'' to convert uncertainties from linear space to logarithmic space: $\err[, log]{x} = \dfrac{\err{x}}{\obs{x} \ln{10}}$ (same for $\err[, log]{y}$). In this approach, we exclude data points where $\obs{x} < 0$ or $\obs{y} < 0$, in accordance with common practice. In contrast, no data points are discarded when using \leopy\ or the two proposed methods.

\subsection{Connection and differences of the proposed methods w.r.t. previous methods} \label{sec:work_compare}

A comparison of the different regression methods is presented in \autoref{tab:method_compare}, in terms of the different treatments of the intrinsic distribution of the independent variable $P(\bx)$, the correlation between uncertainty level and intrinsic value $P(\bx_{err}|\bx)$ and $P(\err{y}|y)$, correlation between the dependent and independent variables $P(y|\bx,\btheta)$, and the optimization objective. In the following, we briefly discuss the connections and differences between the methods we propose and the other methods.

\begin{description}
	\item[OLS and WLS] They can be derived from the \mlmethod\ in case the measurement uncertainty of the independent variable is negligible and $y$ is independent of $\err{y}$. 
	\item[ODR and wODR] The two methods and similar techniques \citep[e.g.][]{Hogg2010, Robotham2015} can be viewed as special cases of the \mlmethod\ with the assumption that the intrinsic data points ($\bx$, $y$) are uniformly distributed along the line (or plane for higher dimensions) defined by the linear correlation. 
	\item[mODR] This is a heuristic approach designed to address deviations from the underlying assumptions of OLS, WLS, ODR, and wODR.  As we will show below (\autoref{fig:general_performance_1}), mODR demonstrates better performance than these four methods. However, its performance does not match the performance of other methods which have a solid mathematical foundation. Due to its low computational cost, mODR could be used to provide a good initial guess for the optimization of more sophisticated methods. 
	\item[\linmix, \roxy\ and \tcpu]  \linmix\ can be considered as a variant of the \mlmethod\ that restricts the correlation between $y$ and $x$ to a linear relationship (note that a log-linear relationship is not linear). It employs the GMM instead of the NF to estimate the intrinsic distribution of $x$ and ignores the potential correlation between the uncertainty level and intrinsic value. While GMM is less powerful than NF, its simplicity allows \linmix\ to integrate the estimation of the intrinsic distribution of $x$ and the parameters of the linear relationship within a unified Bayesian framework. This feature is beneficial for users who wish to explore the degeneracy between these two components. \roxy, a recent JAX-based implementation of similar technique, removes the constraint of linear correlation and offers GPU acceleration and the application of HMC for improved performance. \tcpu, another recent contribution, replaces the Gaussian distribution with Student's t distribution to model or condition intrinsic scatter, thereby increasing robustness against outliers. This improvement can be seamlessly integrated into our \mlmethod\ because our framework allows for any model of intrinsic scatter, although it is not utilized in this paper.
	\item[\ltsfit] Unlike parameter estimation methods, \ltsfit\ serves primarily as a preprocessing technique for data containing significant outliers. It identifies potential outliers, allowing for their removal. The remaining data points can then be used as input for other methods, resulting in improved parameter estimation, as demonstrated in the original paper of \ltsfit\ \citep{ltsfit2013}.
	\item[\leopy] This can be seen as a variant of the \mlmethod\ that takes the intrinsic distribution of $x$ as input, while ignoring the potential correlation between the uncertainty level and intrinsic value. As pointed out in \autoref{sec:dicussion}, the intrinsic distribution of $x$ estimated by the \mlmethod\ and the \ksmethod\ can serve as input for \leopy\ if users wish to explicitly address dependencies among different data points and issues related to censored and missing data.
\end{description} 

\section{Test on log-linear regression} \label{sec:experiment}

\subsection{Mock samples} \label{sec:test_sample}

A log-linear relation can be written as $\log y = k \log x + b$, with an intrinsic scatter of $\sigma$. To perform the regression analyses, we have generated a large set of mock samples by considering wide ranges of $k$, $b$ and $\sigma$, various forms of intrinsic distributions for both $\bx$, complex correlations between variables and their uncertainty level, non-log-linear trends, and the existence of outliers in the data.

For a given set of model parameters ($k$, $b$ and $\sigma$), we first sample $\log x$ from a given distribution $P(\log x)$. We consider 13 different distribution functions for $P(\log x)$, as described in \autoref{app:d_n_setting}. We then compute $\log y$ by $\log y = k \log x + b + \epsilon$, where $\epsilon$ is drawn from a normal distribution with zero mean and variance $\sigma^2$.  Next, for each data point the uncertainty levels $\err{x}$ and $\err{y}$ are sampled according to their correlations with $x$ and $y$, which are also described in \autoref{app:d_n_setting}. Subsequently, we generate noise $e_{x}$ for $x$ and $e_{y}$ for $y$ from Gaussian distributions $N(0, \err{x}^2)$ and $N(0, \err{y}^2)$, respectively. Finally, we calculate the ``observed'' values by $x_{obs} = x + e_{x}$ and $y_{obs} = y + e_{y}$. The log-linear regression analysis is performed basing on $x_{obs}$, $y_{obs}$, $x_{err}$, and $y_{err}$. The model parameters to be determined with the regression methods include the slope $k$, the intercept $b$, and the intrinsic scatter $\sigma$.

\begin{figure*}[ht!]
	\centering
	\includegraphics[width=0.9\textwidth]{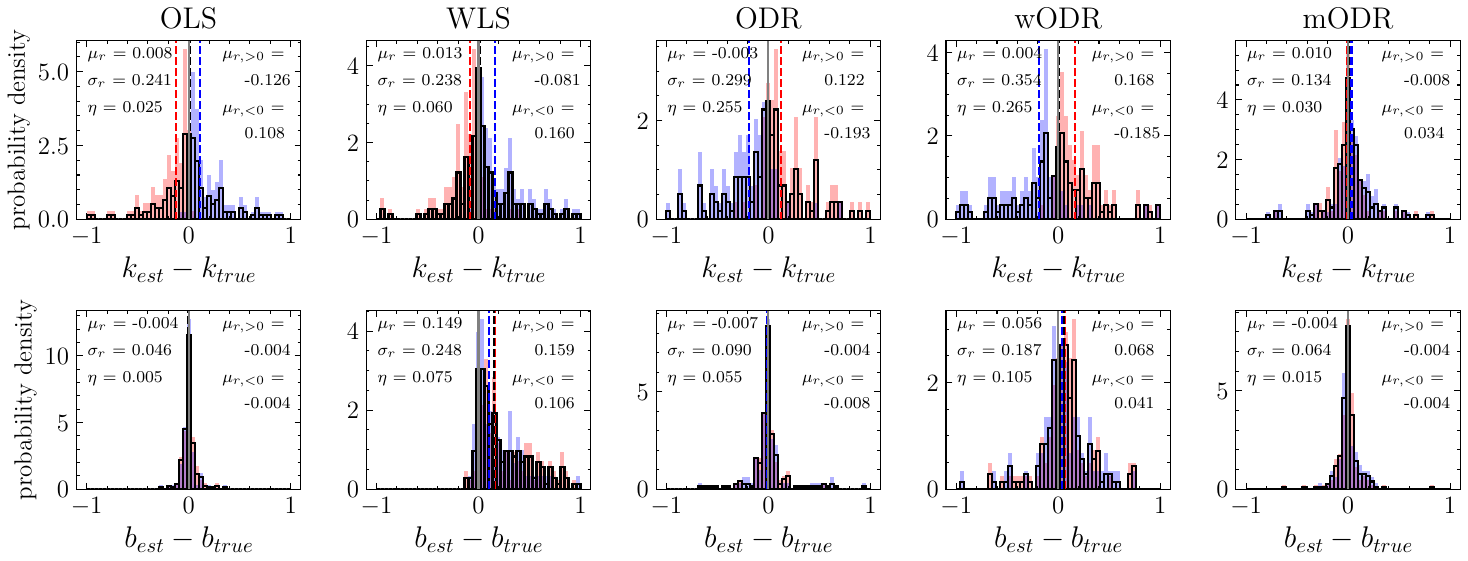}
	\caption{The residual distribution of $k$ and $b$ for OLS, WLS, ODR, wODR, and mODR. The black histogram represents the distribution of all datasets in sample 1, while the red and blue histogram represents the distribution of the datasets with intrinsic value higher and lower than zero, separately. The vertical dashed lines show the median value of the distribution with same color. The median of the residuals for the entire dataset ($\mu_r$), the half of the difference between the 84\% and 16\% quantiles ($\sigma_r$), the extreme error fraction ($\eta$), the median of the residuals for the subsample with true values greater than zero ($\mu_{r, >0}$), and the median for the subsample with true values less than zero ($\mu_{r, <0}$) are presented in each panel. \label{fig:general_performance_1}}
\end{figure*}

\begin{figure*}[t!]
	\centering
	\includegraphics[width=0.9\textwidth]{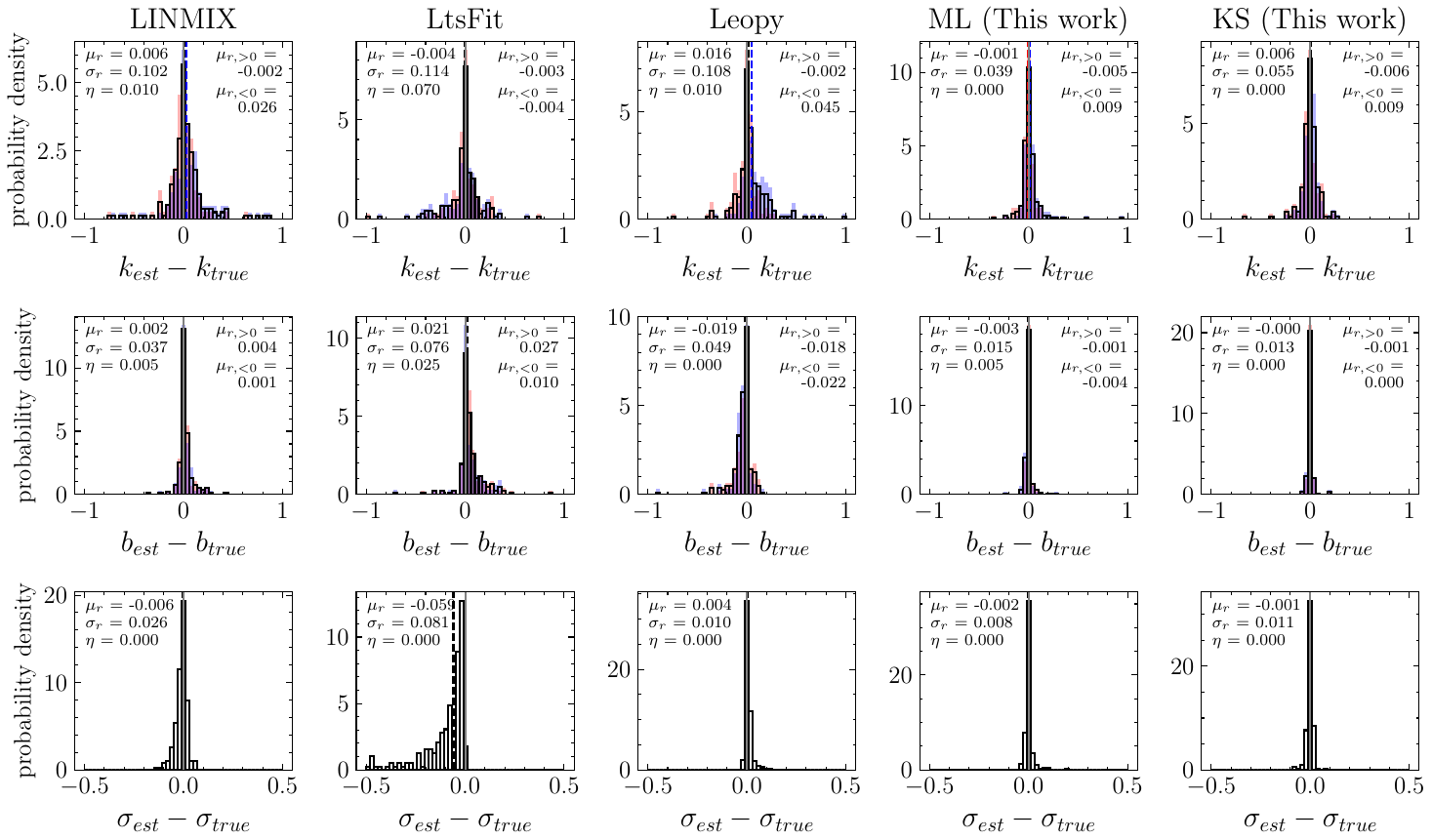}
	\caption{Same as \autoref{fig:general_performance_1} but for \linmix, \ltsfit, \leopy, \mlmethod, and \ksmethod. The third row is added to show the residual distribution of $\sigma$. Only the distribution of all datasets is shown in third row because the non-negative nature of $\sigma$.  \label{fig:general_performance_2}}
\end{figure*}

Accordingly, we have constructed eight mock samples with varying characteristics, designed to evaluate the various aspects of the regression methods. These samples are described below.
\begin{enumerate}
    \item Sample 1 aims to test the general performance of the regression methods. This sample comprises 200 datasets, each with randomly generated slope in the range $-2\leq k \leq 2$, intercept in the range $-2\leq b\leq 2$, and intrinsic scatter in the range $0.1\leq \sigma \leq 0.5$. The distribution of $\log x$ is randomly chosen from ten predefined distributions. The correlation between $x$ ($y$) and $\err{x}$ ($\err{y}$) is randomly selected from three predefined correlations. The typical S/N of $x$ and $y$ are randomly chosen from [1, 10], respectively. Noise is then generated accordingly. The predefined distributions, predefined correlations, and details of noise generation are described in \autoref{app:d_n_setting}.
    
    \item Sample 2 assesses robustness when data exhibit weak non-log-linear trends. It uses the same configuration as the first 40 datasets in Sample 1 (same for other samples below, if not specifically stated). Additionally, a non-log-linear component, formulated as $k k_2 (\log{x})^2 + k k_3 (\log{x})^3$, is added to the basic log-linear relationship between $\log{x}$ and $\log{y}$. The coefficients $k_2$ and $k_3$ are randomly chosen from [-0.3, 0.3].
    
    \item Sample 3 investigates robustness against more complex distributions of $\log x$.  We replace the distribution of $\log x$ with more complex distributions, as described in \autoref{app:d_n_setting}.
    
    \item Sample 4 evaluates robustness when the correlation between $x$ ($y$) and $\err{x}$ ($\err{y}$) is more complex. The more complex correlations used in this sample are described in \autoref{app:d_n_setting}.
    
    \item Sample 5 examines robustness when the estimated $\err{x}$ and $\err{y}$ deviate from their true values. We scale $\err{x}$ and $\err{y}$ for each data point by a factor of 1.1 (Sample 5.1), 0.9 (Sample 5.2), or a random value from [0.9, 1.1] (Sample 5.3), simulating systematic overestimation, systematic underestimation, or uncertainty of $\err{x}$ and $\err{y}$ in real data.
    
    \item Sample 6 assesses robustness in the presence of distribution outliers, which are data points with extreme $x$ values but still following the overall correlation between $\log x$ and $\log y$.  We randomly replace 5\% of the data points by distribution outliers, which are set to be 5–10 times larger/smaller than the maximal/minimal value of $x$. The corresponding $y$ values are calculated based on their replaced $\log x$ values.
    
    \item Sample 7 tests robustness against correlation outliers, where the correlation between $\log x$ and $\log y$ for some data points deviates from the overall correlation. We randomly replace 5\% of the data points with correlation outliers. The $\log x$ and $\log y$ of these outliers have a linear correlation with randomly selected $k$, $b$, and $\sigma$ in the ranges [-2, 2], [-2, 2], and [0.1, 0.5], respectively.
    
    \item Sample 8 assesses robustness when the linear relationship between $\log x$ and $\log y$ becomes a constant value for the largest and smallest parts of $\log x$, a phenomenon we refer to as a ``saturation effect''. We introduce saturation for the data points with the largest and smallest 5\% of $\log x$. Details regarding the saturation effect are presented in \autoref{app:saturate}. 
\end{enumerate}

To maximize the detection of differences between methods and minimize the impact of limited data, we utilize 3,000 data points for each dataset within each sample. This choice is the maximum number of data points permissible given computational time constraints. Typical computational times for each method on a sample with 3,000 data points are presented below\footnote{We use Intel(R) Xeon(R) Gold 6248 CPUs and NVIDIA Tesla V100 GPU for this work. We list the number of CPUs used if the method occupies more than one CPU.}: 0.1 s for OLS, WLS, ODR, and wODR; 0.01 s for mODR; 127 s for \linmix\ (80 CPUs, with posterior); 16 s for \ltsfit; 250 s for \leopy\ (80 CPUs); 10 s (GPU) $\times$ 2 (fitting NF for dependent and independent variables) + 144 s (50 CPUs, with approximate posterior) for \ksmethod; and + 3 s (GPU, with posterior) for \mlmethod. Furthermore, for Sample 1, we generate versions with sample sizes of 1000, 500, 300, 100, and 50 to investigate the performance scaling with sample size in \autoref{sec:sample_size}.

For \leopy, the intrinsic distribution of $x$ is required as input for the regression analysis. As this distribution is typically unknown for real data, we employ a heuristic approach to approximate it rather than directly using the known distribution from data generation. First, we use Tweedie's formula with a Gaussian assumption to ``shrink'' $\obs{x}$ as $x_{\rm shrunk} = (1 - B)\obs{x} + B\ {\rm MEAN}(\obs{x})$, where $B = \err{x}^2 / {\rm VAR}(\obs{x})$.  Next, we fit a beta distribution to $x_{\rm shrunk}$ and use this distribution as input for the \leopy\ method. This approach continues to utilize 3,000 data points. It is important to note that the NF-based method we propose is a better alternative for estimating the intrinsic distribution of $x$. The combination of the NF-based method with \leopy, along with its associated benefits and costs, is further discussed in \autoref{sec:dicussion}. For \linmix, we employ three Gaussian components in this test. Our analysis indicates that incorporating additional components does not improve the performance of this method.

For \mlmethod\ and \ksmethod, we need to estimate the joint distribution of intrinsic values and uncertainty levels, $p(x, \err{x})$, from the observed data $(\obs{x}, \err{x})$. This step also uses 3,000 data points. We employ the NF package \pzflow\ developed by \cite{Crenshaw2024, pzflow}, and customize it for implementing the approach detailed in \autoref{sec:intro_method}. We introduce two techniques to accelerate the convergence of NF training\footnote{These two techniques accelerate convergence but do not affect final performance.}. Firstly, we fix the first layer of the NF as $u = h(x / {\rm MEDIAN}(x)), v = h(\err{x} / {\rm MEDIAN}(\err{x}))$, where $h(x) = \log(x + 1)$ if $x > 0$, and $h(x) = x$ otherwise. Secondly, prior to optimizing \autoref{eq:NF_like}, we pre-train the NF on $(x_{\rm shrunk}, \err{x})$ for 200 epochs. We use the {\tt adam} optimizer with a learning rate of $3 \times 10^{-3}$. We employ the {\tt RollingSplineCoupling} bijector from \pzflow\ with 2 layers and 16 bins. The batch size equals the number of training samples, and training halts if the loss does not decrease for 10 epochs. The fitting of $p(y, \err{y})$ follows the same approach. In the majority of cases ($> 95 \%$), this setting yields an NF with a p-value of the \tdks\ between the generated dataset $(\obs[,\rm gen]{x}, \err[,\rm gen]{x})$ and observed dataset $(\obs{x}, \err{x})$ larger than $3 \times 10^{-3}$. For the cases where the \tdks\ fails, we manually tune the NF hyperparameters to pass the test.

\subsection{Results}


\begin{figure*}[ht!]
	\centering
	\includegraphics[width=0.9\textwidth]{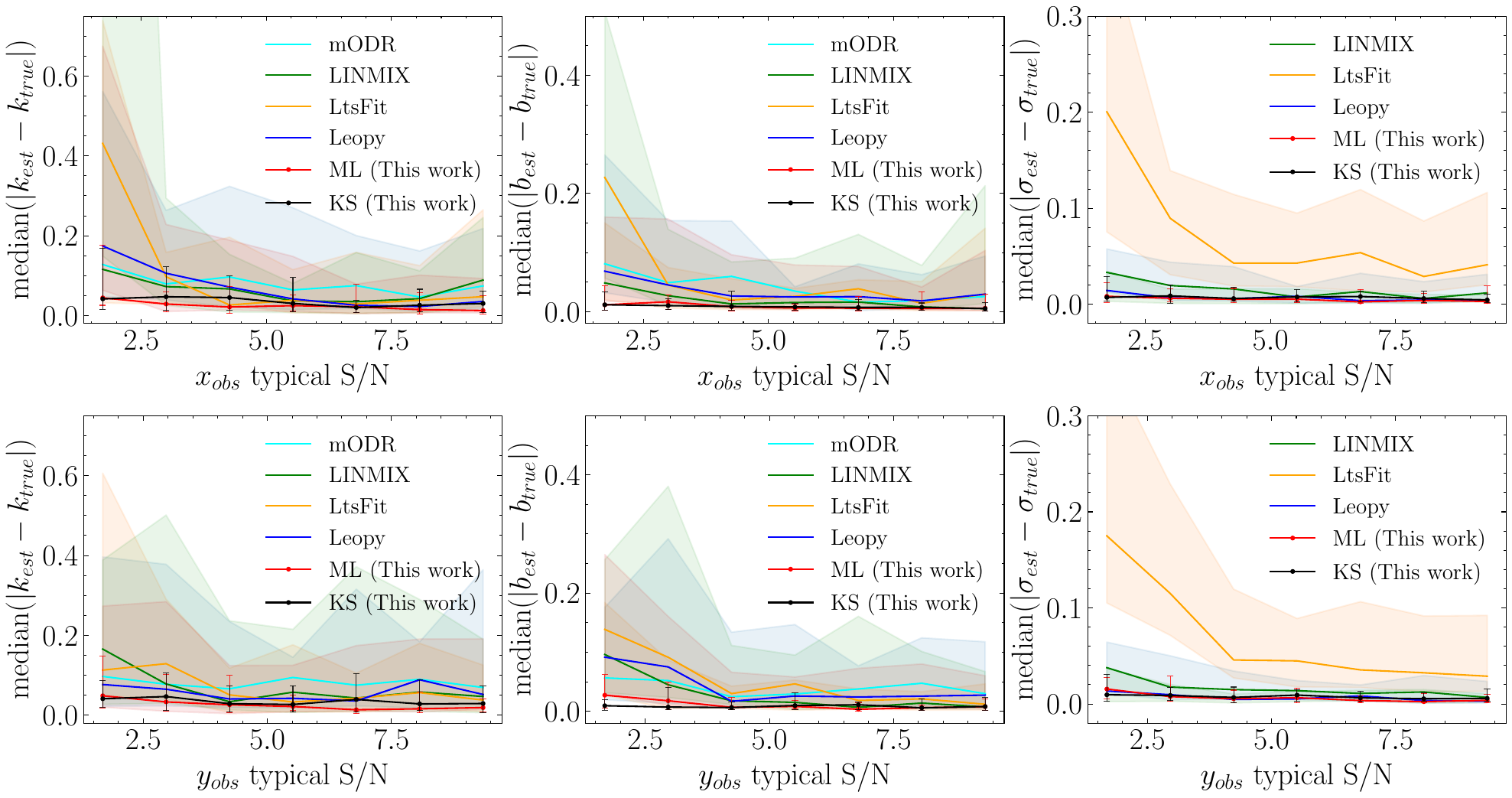}
	\caption{The median absolute error (MAE) of $k$ (left), $b$ (middle) and $\sigma$ (right) as functions of typical S/N of $\obs{x}$ (upper) and $\obs{y}$ (lower), obtained using different regression methods as indicated. The error bars for the \ksmethod\ and \mlmethod\ and the shaded regions for the other methods indicate the range from 25\% quantile to 75\% quantile of the absolute errors within each bin of S/N. \label{fig:scale_with_SNR}}
\end{figure*}

\subsubsection{The overall performance}

The general performance of the methods, evaluated with Sample 1, is presented in \autoref{fig:general_performance_1} in terms of $k$ and $b$ for the OLS, WLS, ODR, wODR, and mODR methods, and in  \autoref{fig:general_performance_2} in terms of $k$, $b$ and $\sigma$ for \linmix, \ltsfit, \leopy, \mlmethod, and \ksmethod.  Each panel displays the distribution of residuals, calculated as the difference between the estimated and true values.  The black histogram represents the overall residual distribution, while the blue and red histograms represent the distributions for subsamples with true values smaller and larger than zero, respectively. Vertical dashed lines in black, blue, and red indicate the medians of the corresponding histograms.  The upper left and upper right corners of each panel provide key performance metrics: the median of all residuals ($\mu_r$); the half of the difference between the 84\% and 16\% quantiles ($\sigma_r$); the extreme error fraction ($\eta$), representing the proportion of residuals with absolute values exceeding 1.0 for $k$ and $b$, or 0.5 for $\sigma$; the median of residuals for the subsample with true values greater than zero ($\mu_{r, >0}$); and the median for the subsample with true values less than zero ($\mu_{r, <0}$).  These metrics, excluding $\eta$, are calculated after clipping extreme residuals. 

As can be seen, mODR, \linmix, \ltsfit, \leopy, \mlmethod, and \ksmethod\ achieve significantly higher accuracy compared to OLS, WLS, ODR, and wODR. The latter four methods also exhibit systematic bias in the estimation of $k$: OLS and WLS tend to overestimate $k$ when $k_{true} < 0$ but underestimate $k$ when $k_{true} > 0$, while ODR and wODR show the opposite trend. Additionally, WLS systematically overestimates $b$. These biases do not indicate fundamental flaws in these methods but arise from the violation of their underlying assumptions in the log-linear regression problem considered here, which is also prevalent in numerous real astronomical datasets. For simplicity, these four methods will not be further considered in the rest of this section. 

The remaining six methods do not exhibit significant bias in the estimation of all parameters, except for the intrinsic scatter estimated by \ltsfit. Notably, the two methods proposed in this work present the best performance, with \ksmethod\ achieving an extreme error fraction ($\eta$) of 0 for all three parameters, highlighting its robustness. 

\subsubsection{Performance across different S/N}

\autoref{fig:scale_with_SNR} displays the Median Absolute Error (MAE), defined as the median of the absolute residuals, as a function of the typical S/N of both the independent (upper panels) and dependent (lower panels) variables in Sample 1. The colored lines and error bars/shaded regions represent the median and scatter resulted from different methods, as indicated. We see a general decrease in MAE with increasing typical S/N. Among others, our proposed methods exhibit the lowest MAE and a flatter trend, demonstrating their effectiveness even at S/N levels around 1. 

\subsubsection{Robustness to the violation of the log-linear assumption}

\begin{figure*}[ht!]
	\centering
	\includegraphics[width=0.9\textwidth]{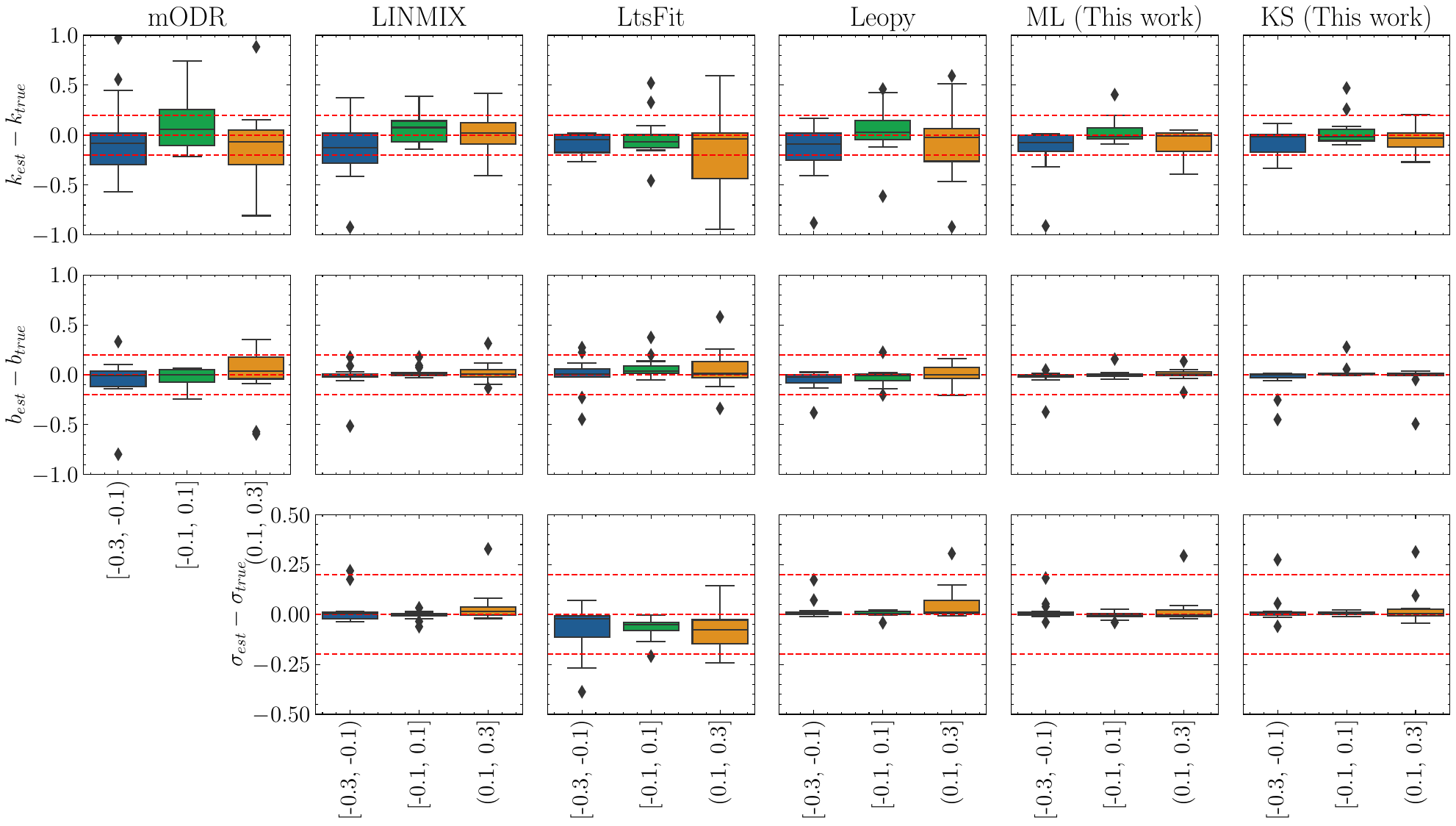}
	\caption{The boxplots of the residual of estimated parameters within different $k_2$ bins in sample 2. A horizontal dashed red line at zero indicates no bias in the residuals. For reference, we include additional horizontal dashed red lines at +0.2 and -0.2. \label{fig:nonlinear_k2}}
\end{figure*}

\begin{figure*}[ht!]
	\centering
	\includegraphics[width=0.9\textwidth]{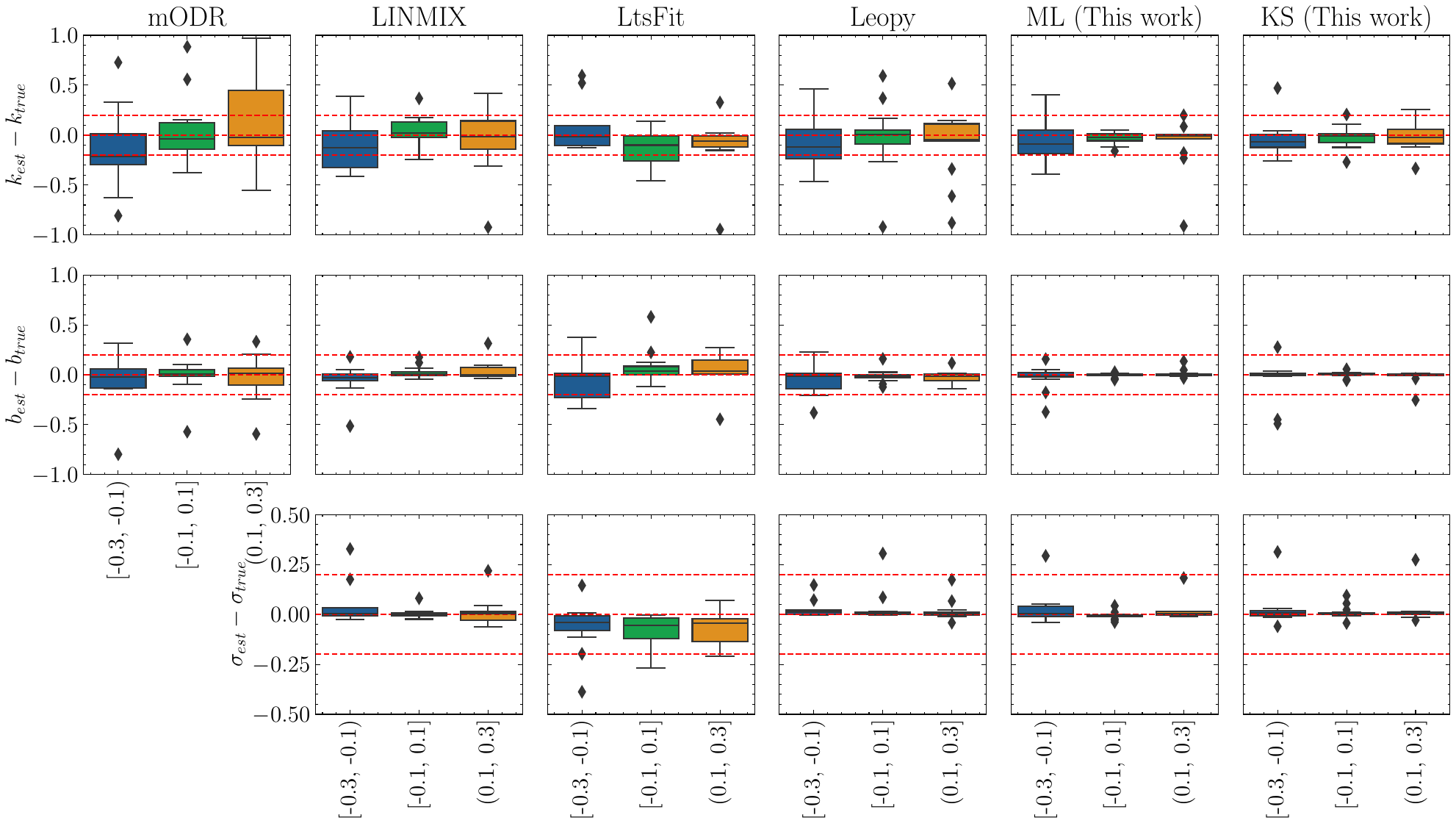}
	\caption{Same as \autoref{fig:nonlinear_k2} but for different $k_3$ bins. \label{fig:nonlinear_k3}}
\end{figure*}

The robustness against violations of the log-linear assumption is evaluated by applying the various methods to Sample 2, where a non-log-linear component formulated as $k k_2 (\log{x})^2 + k k_3 (\log{x})^3$ is additionally included in the log-linear relation. \autoref{fig:nonlinear_k2} displays boxplots of the residuals for different parameters across various $k_2$ bins for each method.  Each boxplot represents the 25\% quantile ($q_{25}$) and the 75\% quantile ($q_{75}$) of the residuals as its edges, with the median value shown as a horizontal line within the box. Data points exceeding $q_{75} + 1.5(q_{75} - q_{25})$ (plotted as the top error bar) or falling below $q_{25} - 1.5(q_{75} - q_{25})$ (plotted as the bottom error bar) are identified as outliers and denoted by black diamonds. \autoref{fig:nonlinear_k3} presents the same analysis, but for different $k_3$ bins. The results demonstrate that the estimation of $k$ is significantly affected by the presence of nonlinear terms. Among the six methods, \ksmethod\ is least affected by nonlinear terms, highlighting its robustness. 

\begin{figure*}[ht!]
	\centering
	\includegraphics[width=0.9\textwidth]{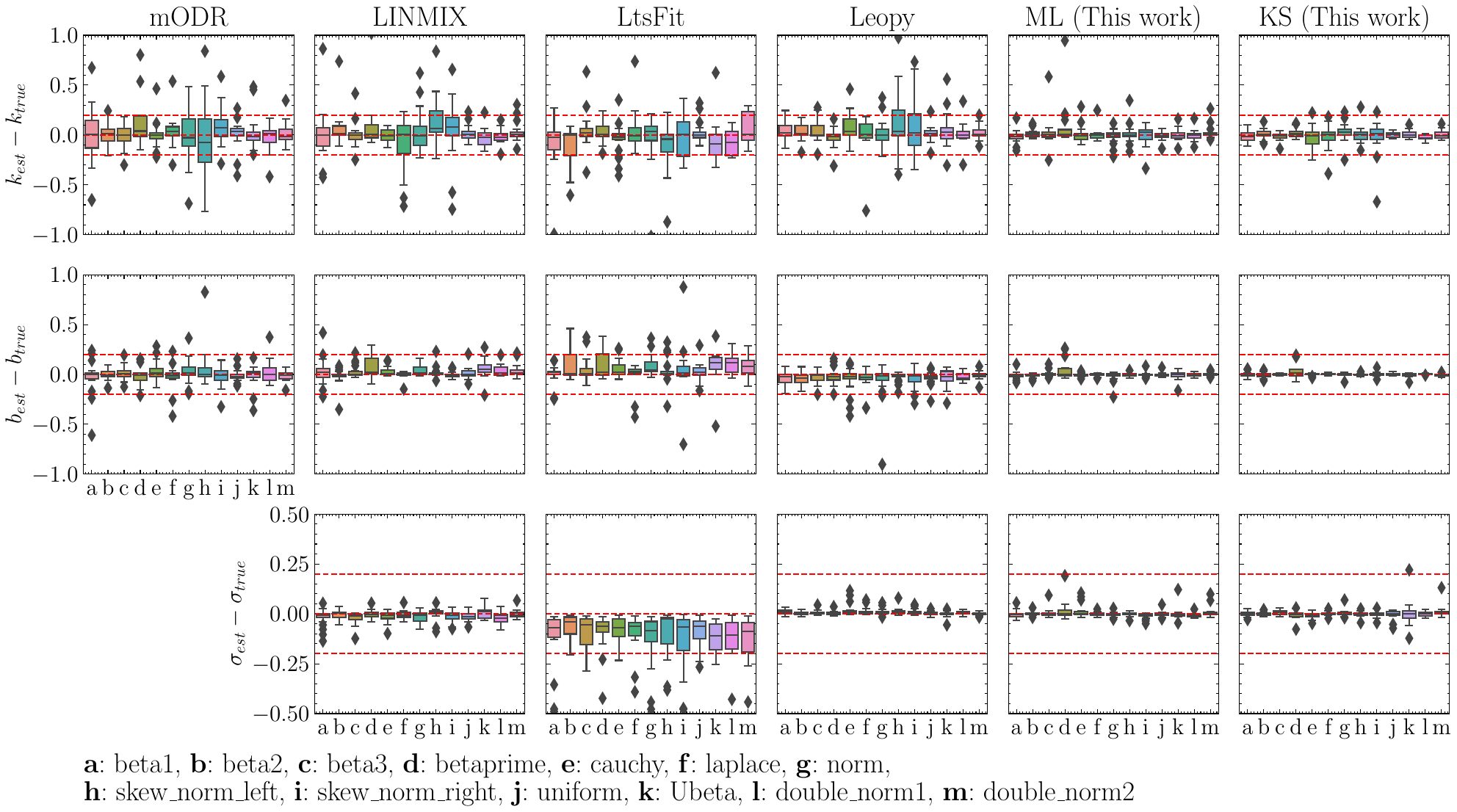}
	\caption{The boxplots of the residual of estimated parameters for datasets from sample 1 and sample 3 with different intrinsic $\log x$ distributions. Horizontal dashed red lines have the same sense as that in \autoref{fig:nonlinear_k2}. \label{fig:diff_sampler_type}}
\end{figure*}

\subsubsection{Robustness to different intrinsic distributions of independent variable}

We use both Sample 1 and Sample 3 to investigate the performance of different methods under various intrinsic distributions of the independent variable. Sample 1 is generated using ten basic distributions, while Sample 3 is generated using three additional complex distributions (see \autoref{app:d_n_setting} for details). The boxplots of the residuals in the parameter estimation are shown in \autoref{fig:diff_sampler_type}. The performance of all methods, except the \mlmethod\ and \ksmethod, is sensitive to the intrinsic distribution of the independent variable. This demonstrates the advantage of using NF to estimate the intrinsic distribution from the dataset.

\subsubsection{Robustness to different correlations between intrinsic value and uncertainty level}

\begin{figure*}[ht!]
	\centering
	\includegraphics[width=0.9\textwidth]{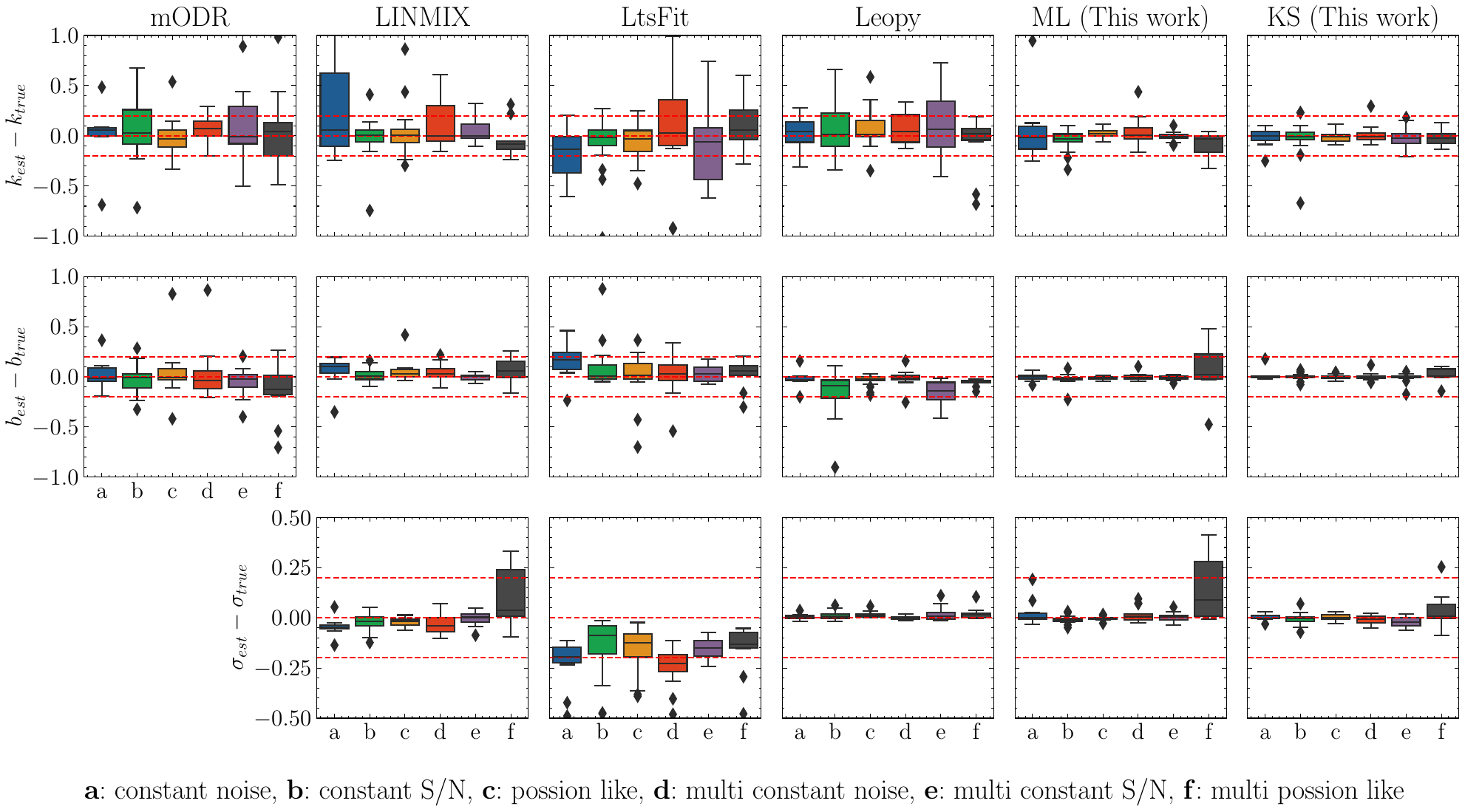}
	\caption{The boxplots of the residual of estimated parameters for datasets from sample 1 and sample 4 with different correlations between intrinsic value and uncertainty level. Horizontal dashed red lines have the same sense as that in \autoref{fig:nonlinear_k2}. \label{fig:diff_error_type}}
\end{figure*}

We use Sample 1 and Sample 4 to evaluate the performance of different methods under varying intrinsic value-uncertainty level correlations (see \autoref{app:d_n_setting} for details). The result is shown in \autoref{fig:diff_error_type}. All the methods except the \ksmethod\ exhibit significant inhomogeneity for the cases of correlations considered. The \ksmethod\ shows nearly homogeneous performance in all the cases, although the estimation of the slope $\sigma$ becomes slightly worse for the ``multi-Poisson-like'' cases. 

\subsubsection{Robustness to the inaccuracy estimation of uncertainty level} \label{sec:inaccuracy_unc}

The uncertainty levels $\err{x}$ and $\err{y}$ are often subject to potential variance and bias in real data.  To evaluate the robustness of different methods to inaccuracies in uncertainty level estimation, we perform analyses using the first 40 datasets from Sample 1 (accurate uncertainty levels), Sample 5.1 (systematically overestimated uncertainty levels), Sample 5.2 (systematically underestimated uncertainty levels), and Sample 5.3 (unbiased uncertainty levels with random noise). Despite the use of inaccurate values for $\err{x}$ and $\err{y}$, the performance of all the methods remains comparable to that obtained with accurate values. This finding suggests that, within the scope of our tests, all the six methods similarly exhibit robustness to both variance and bias in the estimation of $\err{x}$ and $\err{y}$. 

\begin{figure*}[t!]
	\centering
	\includegraphics[width=0.9\textwidth]{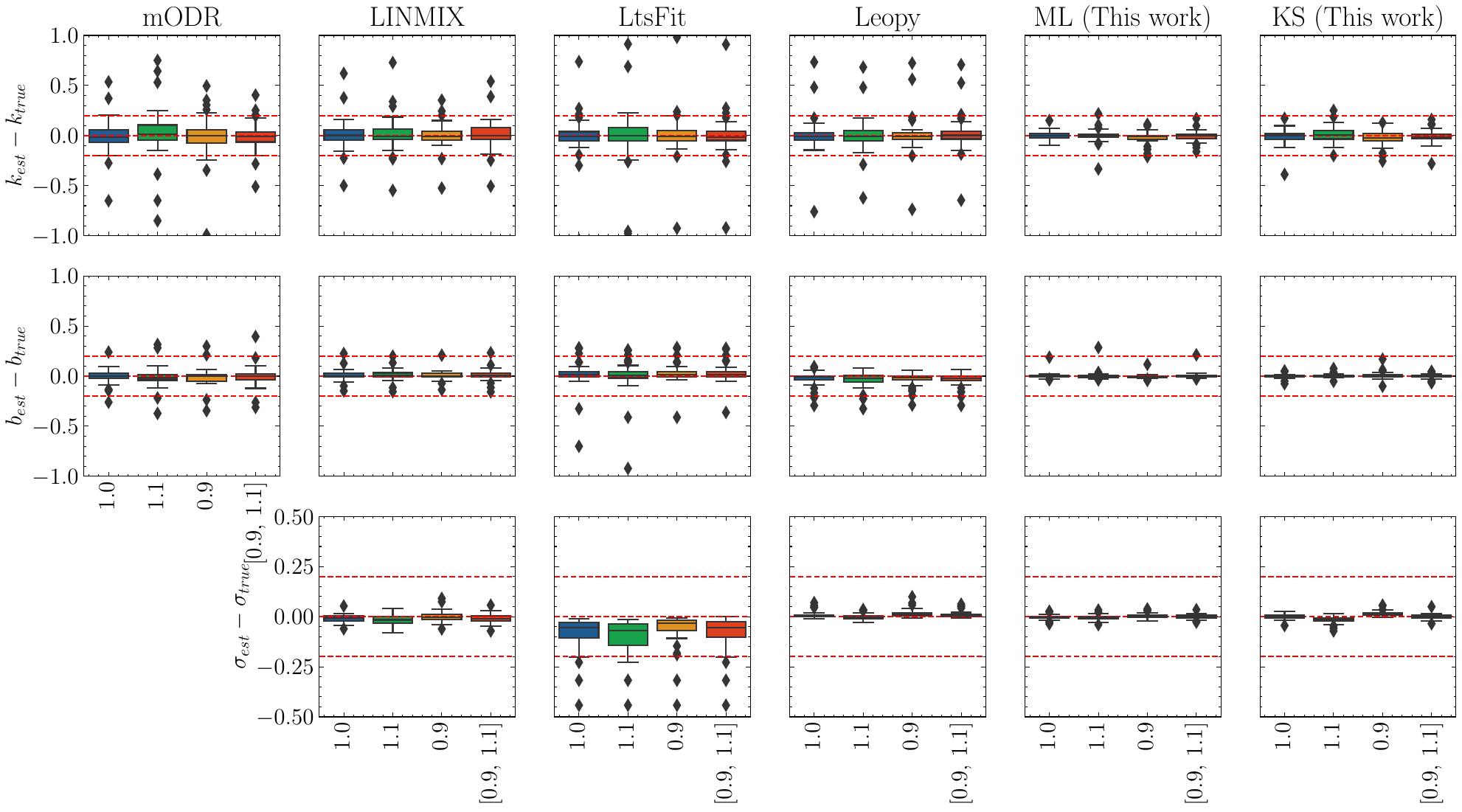}
	\caption{The boxplots of the residual of estimated parameters for datasets with accuracy or inaccuracy observed uncertainty level. In each panel, the first bin contains the first 40 datasets from sample 1, which has accuracy observed uncertainty level. The second and third bins include the dataset from sample 5.1 and 5.2, for which the observed uncertainty level is systematically overestimated and underestimated by a factor of 1.1 and 0.9, separately. The last bin contains the datasets from sample 5.3, in which the observed uncertainty level of each data point is scaled by a random number uniformly distributed in [0.9, 1.1]. Horizontal dashed red lines have the same sense as that in \autoref{fig:nonlinear_k2}. \label{fig:error_of_error}}
\end{figure*}

\begin{figure*}[ht!]
	\centering
	\includegraphics[width=0.8\textwidth]{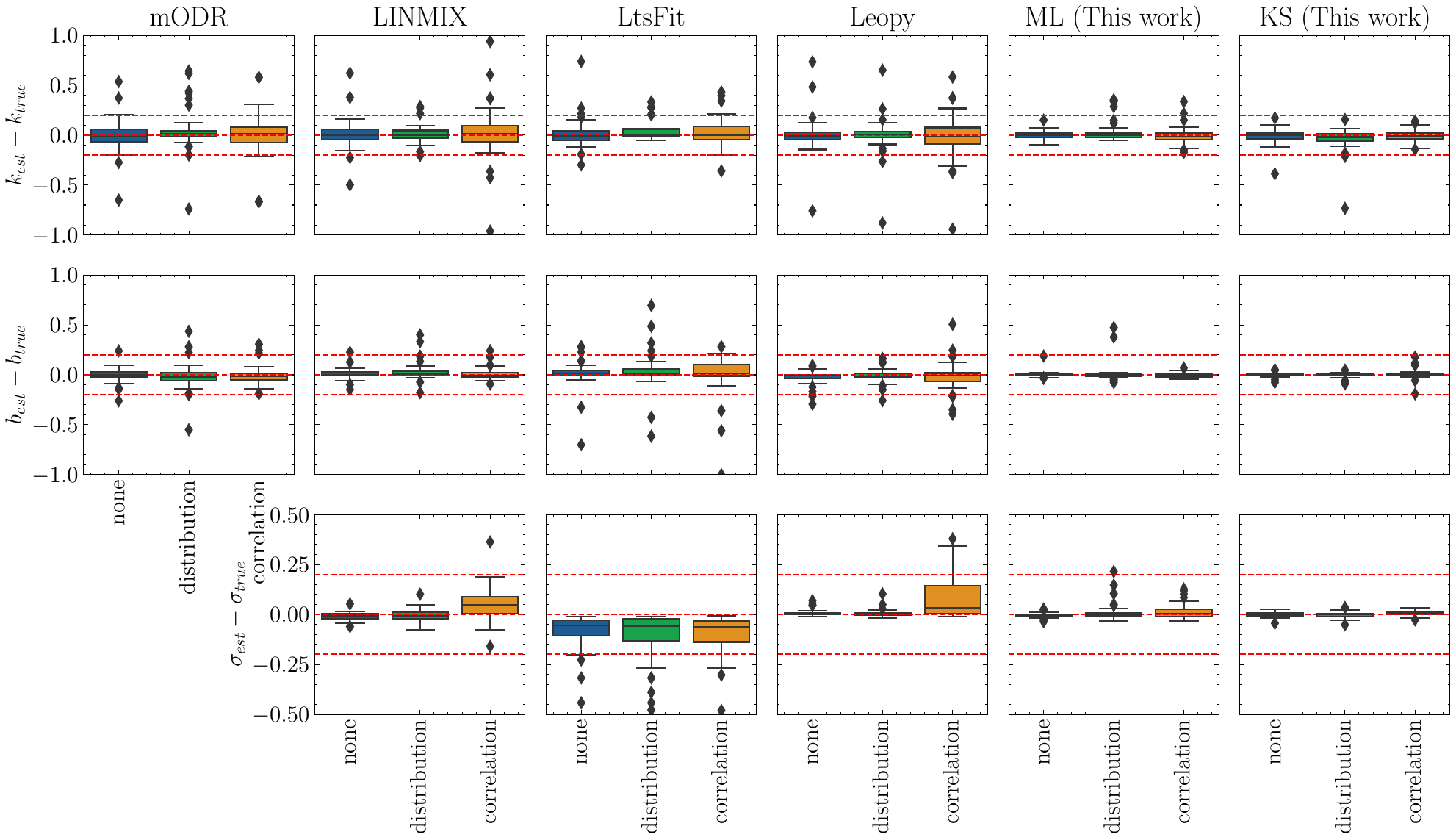}
	\caption{The boxplots of the residual of estimated parameters for datasets without outliers (first 40 datasets from sample 1), with distribution outliers (sample 6), and with correlation outliers (sample 7). Horizontal dashed red lines have the same sense as that in \autoref{fig:nonlinear_k2}. \label{fig:outlier}}
\end{figure*}

\begin{figure*}[ht!]
	\centering
	\includegraphics[width=0.8\textwidth]{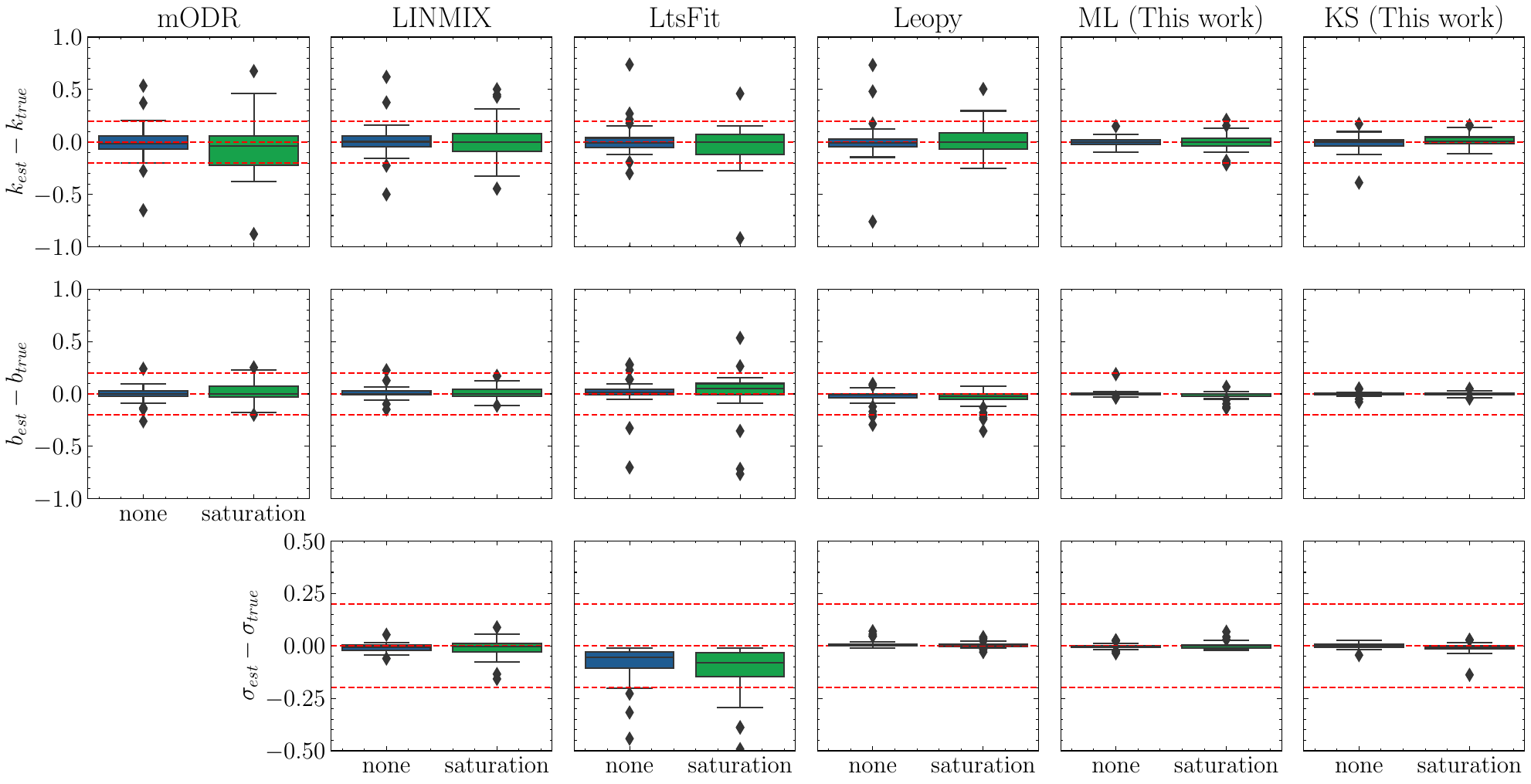}
	\caption{The boxplots of the residual of estimated parameters for datasets without saturation effect (first 40 datasets from sample 1), and datasets with saturation effect (sample 8). Horizontal dashed red lines have the same sense as that in \autoref{fig:nonlinear_k2}. \label{fig:saturation}}
\end{figure*}

\subsubsection{Robustness to outliers}

Real datasets often contain data points with extreme values of $x$ (and consequently $y$), and there is also no guarantee that all data points have the same correlation between $x$ and $y$ as assumed. Such data points with extreme values or deviations from the assumed correlation can be considered outliers. Ideally, a robust regression method should mitigate the influence of these outliers and accurately estimate the parameters for the majority of data that exhibit normal values and conform to the assumed correlation. Here we assess the effect of two different types of outliers as defined in \autoref{sec:test_sample}: the distribution outliers (included in Sample 6) with extreme values of $x$ and consequently $y$, and the correlation outliers (included in Sample 7) deviating from the assumed log-linear relation. \autoref{fig:outlier} shows the residuals of the parameter estimation for the two samples and those of Sample 1 which have no outliers. 

As can be seen, \linmix\ and \leopy\ exhibits lower accuracy with both higher variance and systematic overestimation when estimating the intrinsic scatter $\sigma$ on datasets with correlation outliers. This is because these outliers deviate from the overall correlation exhibited by the majority of data, leading to an apparent increase in scatter. In contrast, both the \mlmethod\ and the \ksmethod\ are less affected by this issue. Furthermore, the estimation of $k$ and $b$ is not significantly influenced by outliers for all the methods tested. One potential solution to account for the effect of outliers is to incorporate an additional component into the model \citep[e.g.][]{Hogg2010, Feldmann19}. However, this approach essentially changes the underlying assumptions, and it is challenging to find a model that effectively describes outliers in real datasets. Our test here shows that the influence of outliers can be minimized by improving the regression algorithm itself, as in the case of our proposed methods. 

\begin{figure}[t!]
	\centering
	\includegraphics[width=0.45\textwidth]{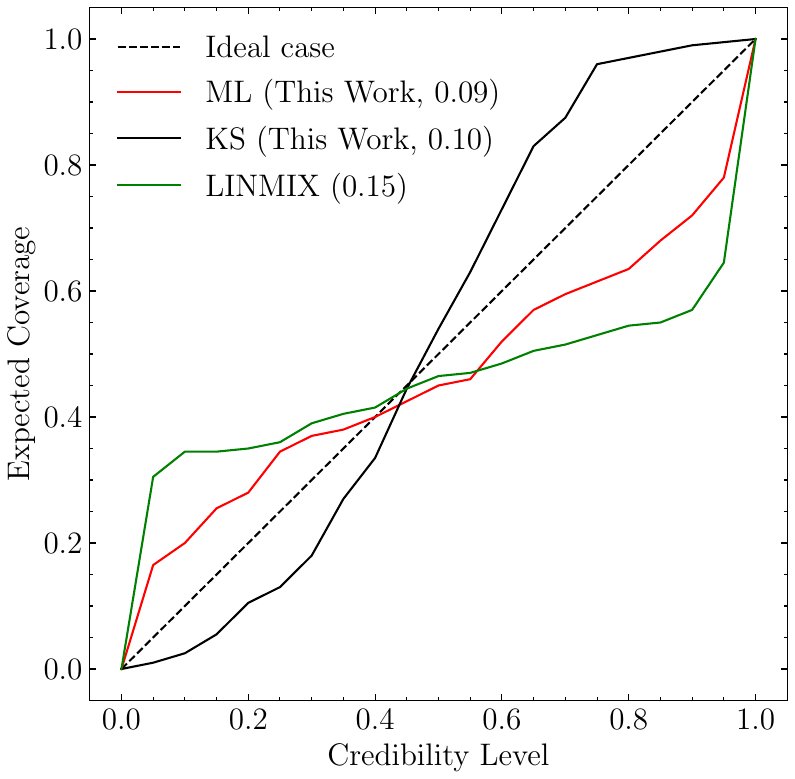}
	\caption{Coverage test for \ksmethod, \mlmethod, and \linmix. \label{fig:err_acc}}
\end{figure}

\subsubsection{Robustness to the saturation effect} \label{sec:saturation}

Commonly seen in real datasets, the saturation effect is a phenomenon where the correlation between $\log y$ and $\log x$ breaks down at one or two ends of the relation, resulting in $\log y$ asymptotically approaching a constant value (see \autoref{app:saturate} for more details). To assess the robustness of the different methods to this effect, we compare the performance on Sample 8 which includes the saturation effect and the first 40 datasets of Sample 1 without  saturation. The result is shown in \autoref{fig:saturation}. The saturation effect significantly impacts the performance of mODR, \linmix, \ltsfit, and \leopy. In contrast, \mlmethod\ and \ksmethod\ exhibits no discernible impact.

\subsubsection{The accuracy of uncertainty estimation} \label{sec:unc_of_unc}

Beyond accurate parameter estimation, quantifying the uncertainty associated with these estimates is essential. In this work, we assess the accuracy of uncertainty estimation by applying a coverage test based on {\tt tarp}\footnote{Official GitHub link: https://github.com/Ciela-Institute/tarp; we use a fork with support for weighted samples in this work: https://github.com/astro-jingtao/tarp} \citep{Lemos2023_tarp} across three methods implemented within a Bayesian or approximate Bayesian framework in our numerical experiments: \linmix, \mlmethod, and \ksmethod. For \linmix, we derive the posterior distribution using a standard MCMC approach. For \mlmethod\ and \ksmethod, we obtain the posterior distribution as described in \autoref{sec:intro_method}.

\autoref{fig:err_acc} presents the coverage test results for the three methods. The results indicate that both \linmix\ and the \mlmethod\ tend to underestimate posterior uncertainties, while the \ksmethod\ tends to overestimate them. We further quantify the degree of inaccuracy using the area between coverage curve of each method and the ideal coverage curve. According to this metric, the \mlmethod\ (0.09) and the \ksmethod\ (0.10) perform better than \linmix\ (0.15). Given that none of the tested methods provide completely unbiased uncertainty estimates, we advise that uncertainties reported by these regression methods should be interpreted cautiously.

\subsubsection{Effect of limited sample size} \label{sec:sample_size}

\begin{figure*}[ht!]
	\centering
	\includegraphics[width=0.8\textwidth]{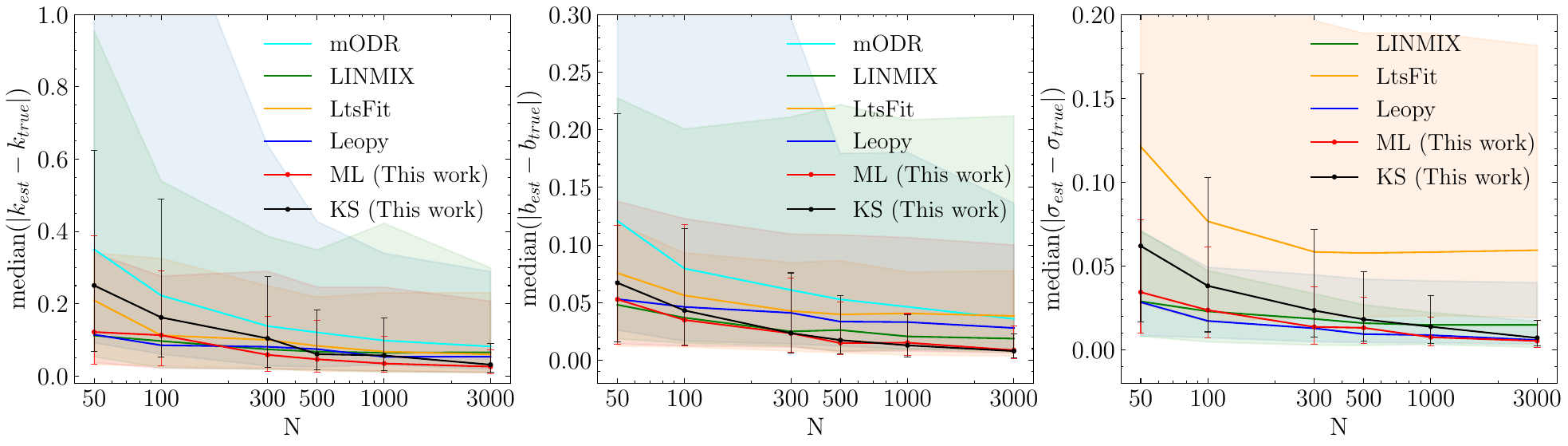}
	\caption{Same as \autoref{fig:scale_with_SNR} but for sample size. \label{fig:diff_N}}
\end{figure*}

The preceding analyses utilize the maximum permissible number of samples given the computational time constraints. However, real-world applications often face limitations in sample size. To assess the effect of sample size, we examine the performance of each method, as depicted in \autoref{fig:diff_N}, using various sample sizes drawn from Sample 1.

We find that, with a sample size ranging from approximately 1000 to 3000, the \mlmethod\ and the \ksmethod\ exhibit comparable performance and outperform other methods, consistent with results shown above. When the sample size is approximately 300–1000, the performance of the \ksmethod\ decreases significantly, becoming comparable to the best-performing of the other methods. In contrast, the \mlmethod\ maintains its advantages. However, with a sample size less than 300, the performance of the \mlmethod\ also becomes comparable to that of the best-performing of the other methods. This test demonstrates that the primary advantage of the proposed methods stems from their ability to efficiently leverage information from large samples, a characteristic expected given that both NF and \tdks\ require substantial sample sizes for optimal performance.

This analysis suggests that method selection should depend on the available sample size. If a sufficient sample size is available ($> 1000$), the \ksmethod\ is preferable due to its high performance and robustness. For sample sizes around 300–1000, the \mlmethod\ is recommended for better performance. Even with limited sample sizes ($< 300$), although the \mlmethod\ exhibits comparable performance to \leopy\ and \linmix, its high computational efficiency, enabled by GPU acceleration, renders it a viable option.

\subsection{Summary of the test} \label{sec:sum_exp}

Our test demonstrates that the \mlmethod\ and the \ksmethod\ exhibit the best and comparable overall performance among other tested methods, provided a sufficient sample size ($> 1000$). Notably, the \ksmethod\ displays greater robustness to the presence of non-log-linear terms, variations in the intrinsic distributions of independent variables, different correlations between intrinsic values and uncertainty levels, inaccuracies in observed uncertainty levels, the existence of distribution and correlation outliers, and the saturation effect. Thus, \ksmethod\ is the preferred choice when a large sample size is available.

Regarding the accuracy of uncertainty estimation, none of the tested methods (\linmix, \mlmethod, and \ksmethod) perfectly capture the uncertainty of all the estimated parameters. The \ksmethod\ tends to overestimate uncertainty, while the \mlmethod\ and \linmix\ tend to underestimate it. Compared to \linmix, \mlmethod\ and \ksmethod\ produces uncertainty estimates closer to the ideal case.

For sample sizes around 300–1000, \mlmethod\ outperforms both \ksmethod\ and the other methods. When the sample size is limited ($< 300$), the performance of \mlmethod, \linmix, and \leopy\ is comparable; however, \mlmethod\ has the advantage of being able to utilize a GPU. Therefore, in these two latter cases, \mlmethod\ is recommended.

\section{Application to astronomical data} \label{sec:real_data}

As a case study, this section applies the regression techniques discussed above to study the correlation of CO(2-1) emission line flux with mid-infrared fluxes down to scales of $\lesssim 100$ pc, using recent observational data from the PHANGS-ALMA \citep{Leroy2021_PHANGS_ALMA_pipeline, Leroy2021_PHANGS_ALMA_survey} and PHANGS-JWST \citep{Lee2023_PHANGS_JWST, Williams2024_PHANGS_JWST} surveys. We consider four nearby galaxies: IC 5332, NGC 0628, NGC 1365, and NGC 7496, for which the observational data was available when we started writing this paper. Similar data have been used for regression analyses of the CO versus MIR flux correlations, e.g. by \citet{Leroy2023} using the mODR method and by \citet{Chown2024} using \linmix. For comparison and for simplicity, here we consider three methods, including both mODR and \linmix, as well as the \ksmethod\ which is shown above to perform best among other methods. In this analysis we focus on the comparison between the three regression methods. Discussion on scientific implications will be presented in a parallel paper (Jing \& Li, in prep.), where we perform a comprehensive application of the \ksmethod\ to the full sample of  19 galaxies in the PHANGS-ALMA and PHANGS-JWST surveys.  

\subsection{PHANGS-ALMA and PHANGS-JWST data}

The data used here include CO(2-1) emission line flux obtained with ALMA, and MIR fluxes at F770W, F1000W, F1130W, and F2100W bands obtained with JWST (version {\tt v0p4}), for four nearby galaxies: IC 5332, NGC 0628, NGC 1365, and NGC 7496. The {\tt v0p4} version of JWST data  doesn't provide uncertainties for the MIR fluxes, which are needed for the regression analysis. We derive the uncertainties as follows. 

First, we compare the {\tt v0p4} data and the original data\footnote{The original data can be found in MAST: \dataset[10.17909/q0wj-xp56]{http://dx.doi.org/10.17909/q0wj-xp56}} available from the Mikulski Archive for Space Telescopes (MAST), and we use the difference between the two datasets  to recalibrate the World Coordinate System (WCS) and the photometry zero point for the original data. The recalibration is determined in such a way that the MAST data and the PHANGS-JWST data are best matched for each band in each galaxy. Then, we apply the corresponding WCS and photometry transformation to the uncertainties in the original data. These transformed uncertainties are considered as the uncertainties of the PHANGS-JWST data. While the transformed MAST flux does not perfectly match the flux in the PHANGS-JWST data, the difference is substantially small, indicating the validation of the recalibration of WCS and zero point as a first-order approximation.  The difference in uncertainties before and after transformation is approximately 10\%, which can be considered the upper limit of the difference resulting from higher-order terms. Based on our test in \autoref{sec:inaccuracy_unc}, this level of higher-order difference does not significantly affect the performance of the regression techniques. The PHANGS-JWST at different bands and the PHANGS-ALMA data are matched through Gaussian convolution to have the same point spread function (PSF). Finally, the uncertainty of the PSF-matched data is derived by error propagation, taking into account the correlation between nearby pixels following \cite{Klein2021}. 

Our method relies on the assumption that data points are independent (see \autoref{sec:dicussion} for further discussion). To address the correlation between pixels, we initially downsample the pixel size to half of the PSF size, as measured by the Full Width at Half Maximum (FWHM). Subsequently, we select a random subsample of 10,000 data points for regression analysis. The data points in this subsample are anticipated to be approximately independent. We observe that different randomly selected subsamples yield highly similar results.

\subsection{Log-linear regression analysis}

\begin{figure}[t!]
	\centering
	\includegraphics[width=0.4\textwidth]{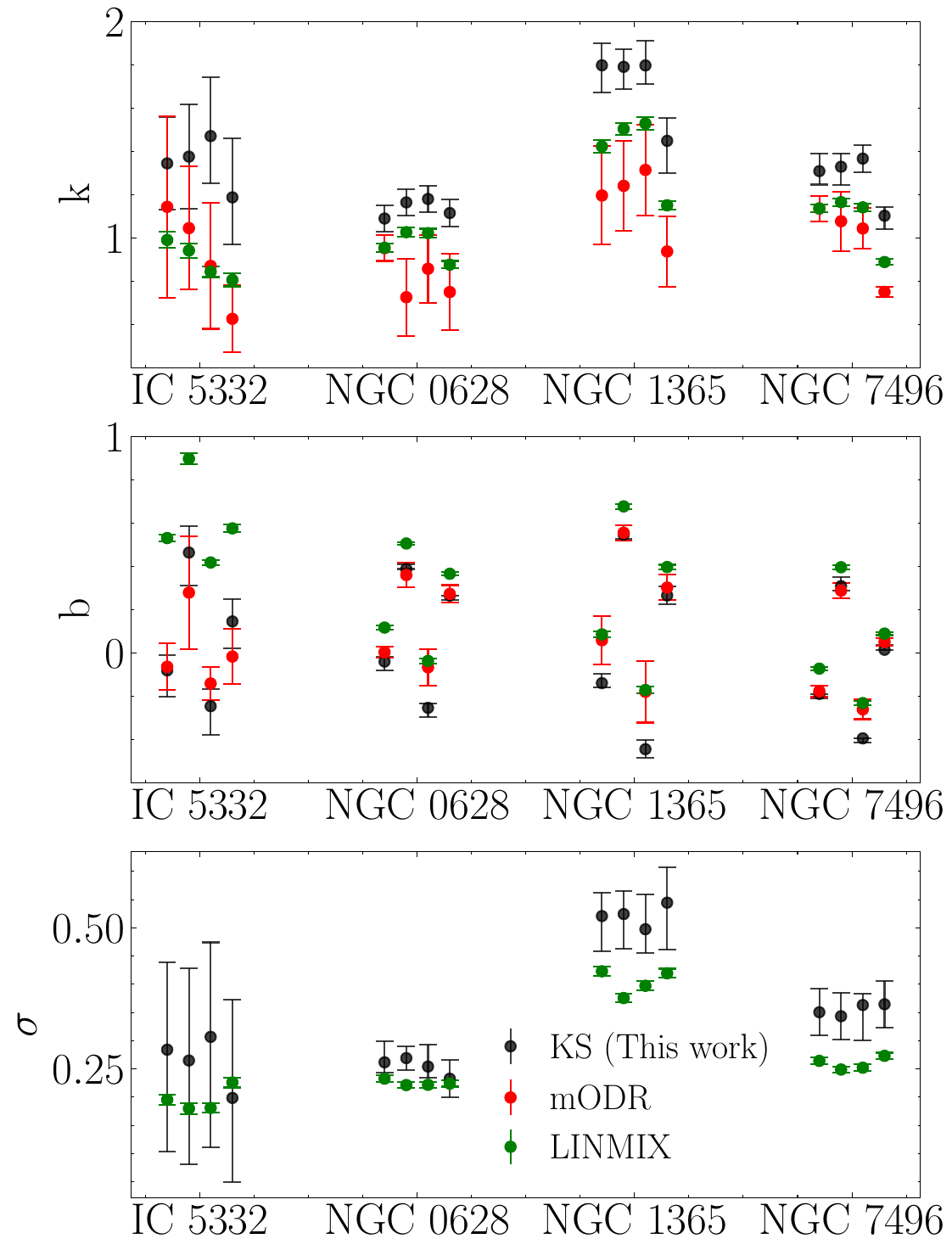}
	\caption{The best-fitting parameters $k$, $b$ and $\sigma$ (alongside their 1$\sigma$ uncertainties represented as error bars) of the log-linear correlation between the CO(2-1) flux and the mid-infrared fluxes at four JWST bands (F770W, F1000W, F1130W, and F2100W, ordered from left to right for a given galaxy), as obtained for the four galaxies (as indicated) and using three regression methods: the \ksmethod\ (black dots), mODR (red dots) and \linmix\ (green dots). \label{fig:compare_CO_JWST}}
\end{figure}

For each galaxy, the \ksmethod, mODR, and \linmix\ are applied to derive the slope $k$, intercept $b$, and intrinsic scatter $\sigma$ (except for mODR) for the pixel-level log-linear correlation between the CO(2-1) emission line flux ($I_{\rm CO}$), and the JWST F770W, F1000W, F1130W, and F2100W fluxes ($I_{\rm F770W}$, $I_{\rm F1000W}$, $I_{\rm F1130W}$, $I_{\rm F2100W}$). In \autoref{fig:compare_CO_JWST}, the best-fitting parameters from the three methods are plotted as black, red and green dots with error bars, for different bands across various galaxies. The uncertainty of the mODR method is determined by the standard deviation of results obtained using different numbers of bins and varying lower and upper limits of the independent variable. The uncertainties associated with the \ksmethod\ and \linmix\ are calculated using the methods described in \autoref{sec:unc_of_unc}. As can be seen from the figure, compared to the \ksmethod, both mODR and \linmix\ yield smaller values of $k$, while \linmix\ yields larger $b$ and smaller $\sigma$. The discrepancy in $k$ should be primarily attributed to three factors: saturation effects in the real data, noise of the independent variable, and the difference between log-linear and linear assumptions. Moreover, the higher values of $b$ obtained using \linmix\ likely stem from the degeneracy between $k$ and $b$, and the exclusion of data points with negative flux measurements. 

\begin{figure}[t!]
	\centering
	\includegraphics[width=0.48\textwidth]{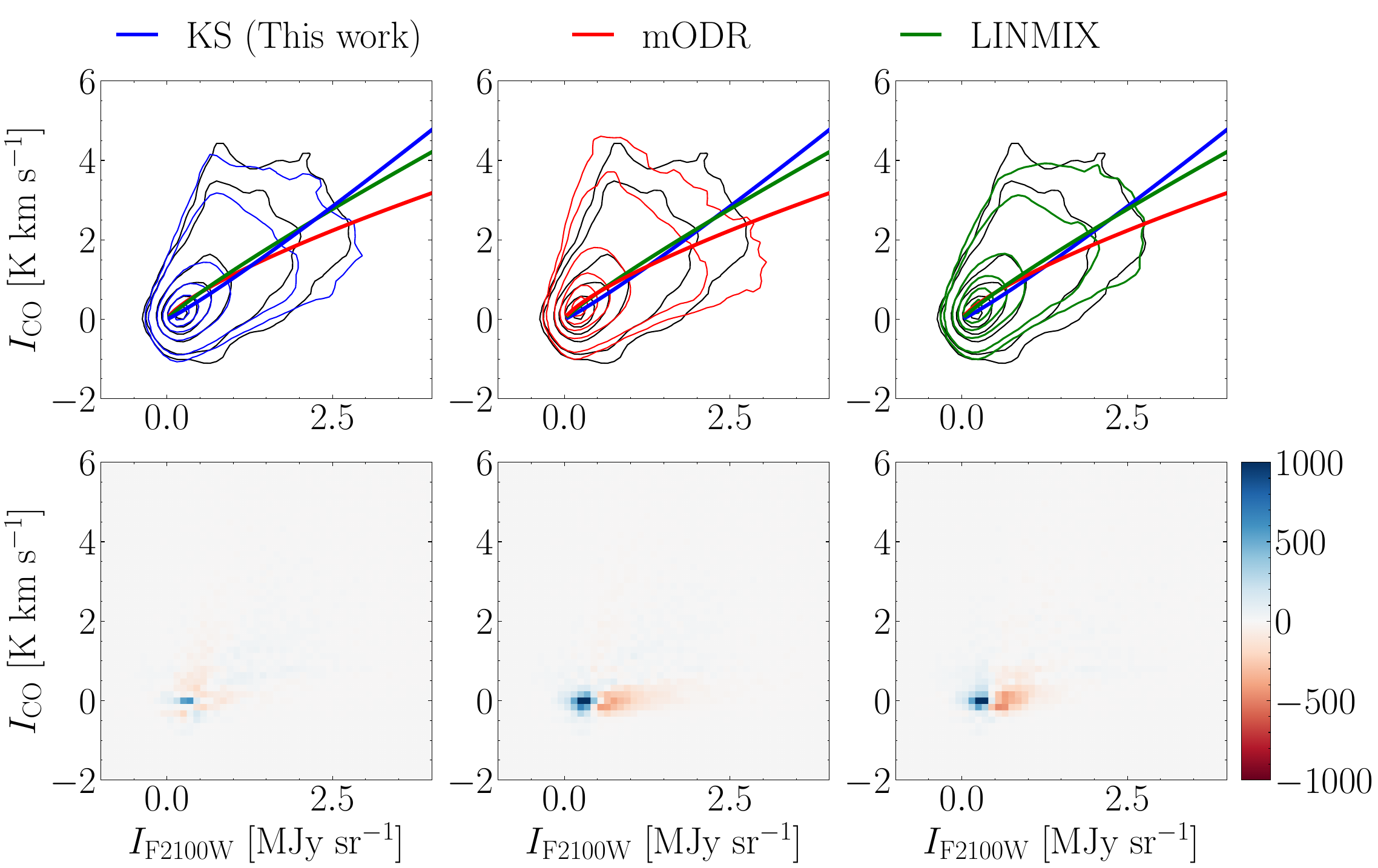}
	\caption{In the first row, the distribution of all data points are shown as the black contours. The blue, red, and green contours represent the distribution of mock data points generated based on the best-fitting log-linear relationship obtained by \ksmethod, mODR, and \linmix, separately. The blue, red, and green lines represent the underlying relationship. In the second row, the differences in the number of data points in each bin between the generated data and the real data ($N_{\rm real} - N_{\rm generated}$) are color-coded. \label{fig:ngc7496_2100w_co21}}
\end{figure}

To better understand the effect of saturation, we generate mock data without saturation for each galaxy based on the best-fitting parameters $k$, $b$, and $\sigma$ derived from each method. The required intrinsic distribution of the independent variable $P(x)$, along with the correlation between uncertainty levels and intrinsic values, $P(\err{x}|x)$ and $P(\err{y}|y)$, is derived from \ksmethod. Since the $\sigma$ parameter is not estimated by the mODR method, we explore a range of $\sigma$ values and select the one that minimizes the distribution distance between the generated and observed data. For example, the upper panels of \autoref{fig:ngc7496_2100w_co21} illustrate the correlation between $I_{\rm CO}$ and $I_{\rm F2100W}$ for NGC 7496 in both the real data (black contours, repeated in every panel) and the generated data (blue, red, and green contours in the panels from left to right for \ksmethod, mODR, and \linmix, respectively). The underlying relations estimated by the three methods are plotted in all panels as the blue, red, and green lines. The lower panels show the differences between the generated data and the real data. As expected, although various regression methods yield residuals to differing degrees, \ksmethod\ presents the smallest residuals.

\begin{figure}[t!]
	\centering
	\includegraphics[width=0.48\textwidth]{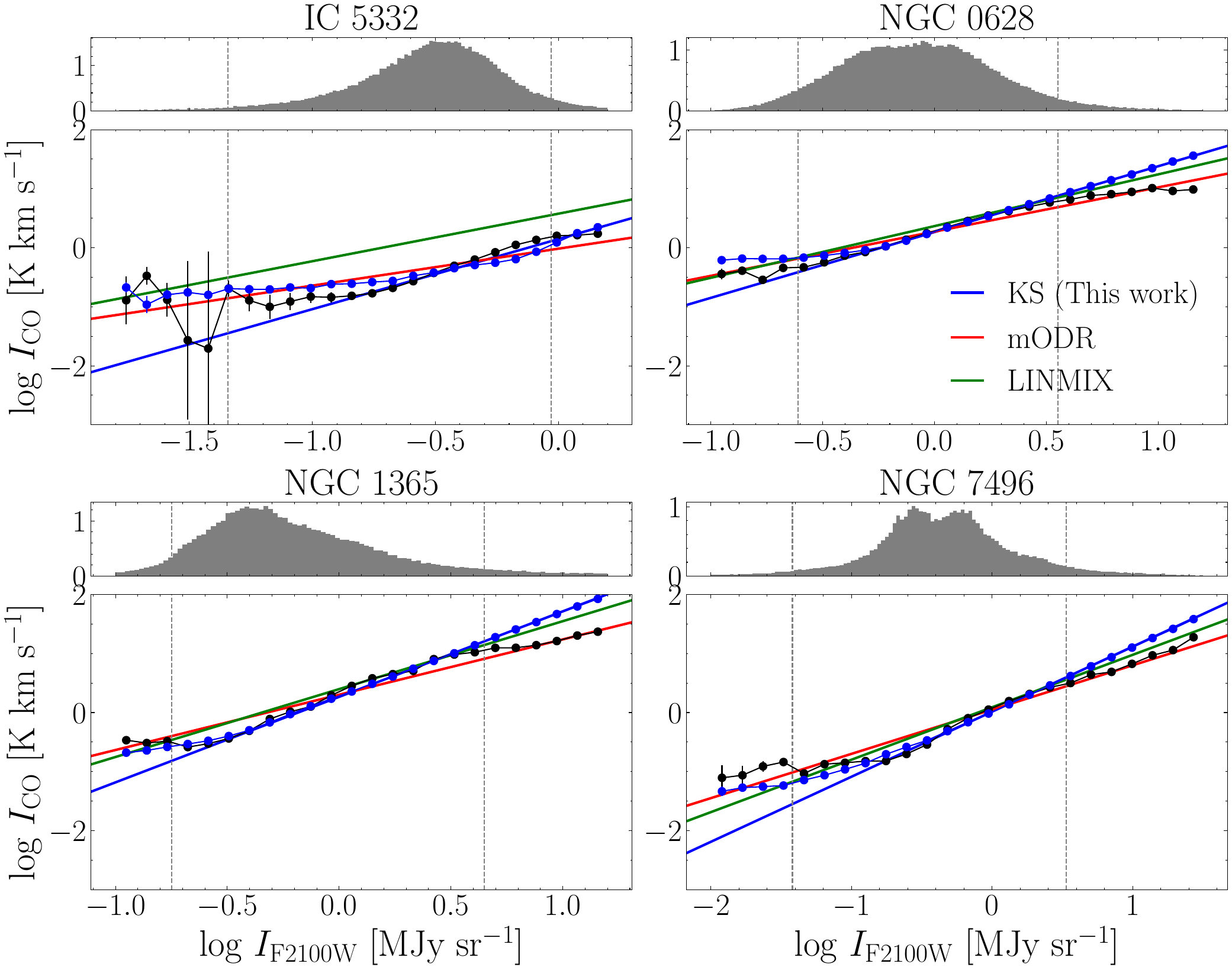}
	\caption{The logarithm of median CO(2-1) flux in each F2100W flux bin for each galaxy is shown as black dots with error bar (for real data, error bar represents 1$\sigma$ uncertainty of median) and blue dots (for mock data generated by the best-fitting results of \ksmethod). The best-fitting relation of \ksmethod, mODR, and \linmix\ on real data are shown as blue, red, and green lines. The two vertical dotted lines in each panel indicate the 5\% and 95\% percentiles of the F2100W flux distribution which is shown in each upper sub-panels. Note that the data points with negative F2100W flux measurements are discarded when plot the marginalized distribution in logarithm space, while included when calculate 5\% and 95\% percentiles. \label{fig:2100w_co21_logspace}}
\end{figure}

The saturation effect is demonstrated in \autoref{fig:2100w_co21_logspace}, which displays the median relation between $\log I_{\rm CO}$ and $\log I_{\rm F2100W}$ for four galaxies. The figure presents the relation in both real (black dots), and generated (blue dots) data basing on \ksmethod. The best-fitting relation obtained by \ksmethod, mODR, and \linmix\ on the real data are shown as blue, red, and green lines, respectively. The two vertical dotted lines indicate the 5\% and 95\% percentiles of the F2100W flux distribution which is shown in each upper sub-panels. For all the galaxies, the generated data match the real data very well at all fluxes below the 95\% percentile, indicative of no saturation effect in the majority of the real data. At the highest fluxes (beyond the 95\% percentile), the relation in the real data falls below that of the generated data, reflecting a significant saturation effect. The best-fitting relation from the \ksmethod\ (the blue line) continuously aligns with the blue dots, even at the high-flux end. This demonstrates that the regression by \ksmethod\ is not biased by the saturation effect, as expected. In contrast, the relations from both mODR and \linmix\ deviate from the blue dots in the saturation-dominated flux range, resulting in flatter slopes (smaller $k$) and larger intercepts (larger $b$), consistent with what is seen above in \autoref{fig:compare_CO_JWST}. 

\begin{figure}[t!]
	\includegraphics[width=0.47\textwidth]{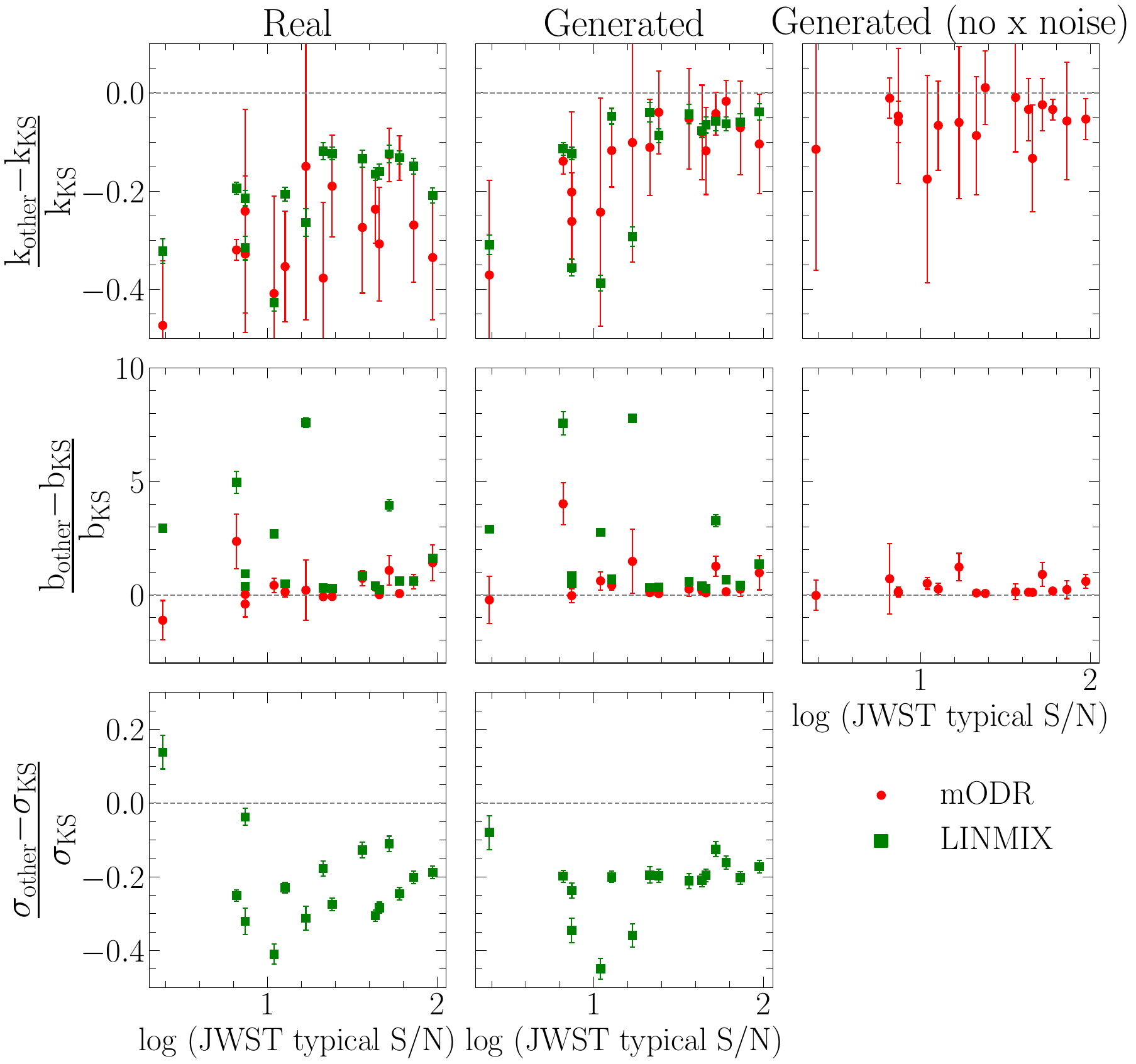}
	\caption{The difference in the estimated k (first row), b (second row), and $\sigma$ (third row) by mODR (red dots) and \linmix\ (green squares) relative to the estimates obtained from \ksmethod\ is shown as a function of the typical S/N of the independent variable. The error bars indicate 1$\sigma$ uncertainties. This analysis is applied on real data (left column), generated data based on the results of \ksmethod\ with noise (center column), and generated data without noise (right column).  The \linmix\ is not applied on generated data without the noise, due to the numerical issue caused by the zero uncertainty. \label{fig:diff_vs_SNR}}
\end{figure}

At low fluxes, a similar flattening trend is seen in the relations of both real data and generated data. This flattening is not attributed to saturation effects, as the generated data are devoid of saturation. Instead, this phenomenon primarily arises from the fact that for data points with the lowest $I_{\rm F2100W}$, the measured values are dominated by random noise, which smears out the correlation between $I_{\rm CO}$ and $I_{\rm F2100W}$. To examine the effect of noise, we generate another set of mock data basing on \ksmethod\ in the same way as above, but excluding the noise in the independent variable (i.e. MIR fluxes at the four JWST bands). We perform the three regression methods to both sets of mock data. In \autoref{fig:diff_vs_SNR}, we show the relative difference of $k$, $b$ and $\sigma$ obtained from mODR (red dots) and \linmix\ (green squares) withe respect to that from the \ksmethod, as a function of the typical S/N of the independent variable. The results are shown for the real data of all galaxies at all JWST bands (left column), and for the generated data with (middle column) and without (right column) including the noise of the independent variable. Due to numerical issues arising from zero uncertainty, \linmix\ is not applied to generated data without noise. 

The figure shows that, generally, the estimation biases in all the three parameters increase as one goes from high to low S/N. When saturation effects are removed (panels in the middle column), the biases are reduced at all S/Ns and almost disappear at the highest S/N, consistent with the above conclusion that the high-flux end is biased by the saturation effect. When the noise of the independent variable is further removed for the mODR method, the bias is significantly reduced for $k$ and largely disappears for $b$, with no obvious dependence on S/N anymore. This result demonstrates that the mODR method is indeed biased by the noise of the independent variable in real data, which is not included in the method. This residual biases, approximately 0.1 in $k$ and $<1$ in $b$ as seen for mODR from the right column, stem from a lack of modeling of the intrinsic distribution of the independent variable as well as a lack of consideration for the correlation between the uncertainty level and the intrinsic value. 

For \linmix, as the noise of the independent variable is already taken into account in the method, the residual biases seen in the middle column of \autoref{fig:diff_vs_SNR} should be attributed to the fact that \linmix\ does not natively handle log-linear regression. \linmix\ assumes both the linear correlation and Gaussian noise occur in linear space, whereas the actual relationship involves a linear correlation in log space and Gaussian noise in linear space. The ``delta method'' employed to transform uncertainty from linear to log space is effective only for data of substantially high S/N. 

In summary, the application of the three regression methods to the real astronomical data showcases the impact of the various effects on the log-linear regression, as discussed above in \autoref{sec:experiment}. In particular, the significant biases caused by the use of mODR and \linmix\ highlight the requirement of new techniques such as the \ksmethod\ proposed in this work to be applied in future work. In a parallel paper as mentioned, we use the new method to comprehensively investigate the CO versus MIR flux correlations again, but for all the 19 galaxies in the PHANGS-ALMA and PHANGS-JWST surveys. 

\section{Discussion} \label{sec:dicussion}


While the application to the test data presented in \autoref{sec:experiment} assumes independent noise in $x$ and $y$, the derivation of our method in \autoref{eq:integral} accounts for the case of dependent noise in both independent and dependent variables. Instead of calculating the likelihood as $P(\yobsi|y, \yerri)P(\bxobsi|\bx, \bxerri)$, one can directly compute $P(\yobsi, \bxobsi|\yerri, \bxerri, y, \bx, r)$, where $r$ represents an additional parameter describing the correlation between the noise in $x$ and $y$.  Similarly, incorporating this dependence in mock data generation for the \ksmethod\ is straightforward. 

A more intricate case arises when data points exhibit correlation with each other.  In principle, this case can be addressed by directly calculating $P(\{D_i\}|\btheta)$ for all data points, explicitly incorporating their correlation. However, this calculation significantly increases computational complexity.  Recognizing that highly correlated data points provide limited additional information, we propose a practical approach involving downsampling the dataset to construct a nearly independent subset for regression analysis.  

The proposed methods are not natively equipped to handle censored (typically upper or lower limits in astronomical data) or missing data.  Key challenges arise from the difficulty of applying NF-based methods to datasets with censored or missing data for estimating the intrinsic distribution of independent variables and the correlation between uncertainty level and intrinsic value, as well as the difficulty of constructing a generative model for such datasets when using the \ksmethod.  

In practical situations where censored data originates from limited S/N, such as data points with S/N $<$ 3 being considered non-detections and treated as censored data with a 3$\sigma$ upper limit, we propose directly utilizing these low S/N data points as normal data points in regression analysis. The effectiveness of our proposed method in producing accurate estimations on low S/N datasets has been demonstrated in \autoref{fig:scale_with_SNR}. However, if missing or censored data stem from other sources and are crucial for regression analysis, more customized modifications to the proposed method, such as explicitly modeling data censoring or missing process, are necessary.

An alternative approach involves leveraging the intrinsic distribution estimated by the NF-based method as input for \leopy, which can handle dependencies between different data points and include censored and missing data, provided the NF-based estimation is not significantly affected by these complexities. However, this approach entails the limitation of not seamlessly incorporating the correlation between uncertainty levels and intrinsic values, and the \ksmethod\ cannot be applied.  Further development of the proposed methods to address more complex correlation behaviors and censored or missing data warrants exploration in future work.

\section{Summary} \label{sec:summary}

We introduce a new regression technique (\autoref{sec:intro_method}), the \mlmethod, and its variant, the \ksmethod, designed to obtain unbiased regression results for typical astronomical data which accommodate uncertainty in both independent and dependent variables, unknown intrinsic distributions of the independent variable, unknown correlation between uncertainty levels and intrinsic values for both independent and dependent variables, as well as arbitrary intrinsic correlation (including but not limited to non-linear scaling relation and intrinsic scatter). The variational inference based empirical Bayes analysis is utilized to estimate the intrinsic distribution as well as correlation between uncertainty levels and intrinsic values from dataset. A machine learning technique of normalizing flow (NF) is employed to model distribution and correlation mentioned above in this approach (\autoref{sec:intro_method}). By leveraging these estimated features, our method comprehensively accounts for the uncertainties in both independent and dependent variables. Compared to the \mlmethod\ which is a likelihood or posterior based approach, the \ksmethod\ based on the \tdks\ further enhances the robustness, guided by the principle of minimal distribution difference. 

We evaluate our methods through tests on both mock data (\autoref{sec:experiment}) with a log-linear relationship (distinct from a linear relationship, as we have shown), and real astronomical data from PHANGS-ALMA and PHANGS-JWST (\autoref{sec:real_data}), and we make comparisons with the results obtained from applying other widely-used methods. 

The tests demonstrate that, if sufficient sample size ($> 1000$) provided, both the \mlmethod\ and the \ksmethod\ consistently achieve better overall performance compared to previous methods (\autoref{fig:general_performance_1}, \autoref{fig:general_performance_2}, \autoref{fig:compare_CO_JWST}), particularly in low S/N scenarios (\autoref{fig:scale_with_SNR}, \autoref{fig:diff_vs_SNR}). The \ksmethod\ exhibits robustness against deviations from the log-linear assumption (\autoref{fig:nonlinear_k2}, \autoref{fig:nonlinear_k3}), complex intrinsic distributions of the independent variable (\autoref{fig:diff_sampler_type}), complex correlations between uncertainty levels and intrinsic values (\autoref{fig:diff_error_type}), inaccurate estimation of uncertainty levels (\autoref{fig:error_of_error}), the presence of outliers (\autoref{fig:outlier}), and the presence of saturation effects (\autoref{fig:saturation}, \autoref{fig:2100w_co21_logspace}, \autoref{fig:diff_vs_SNR}). Furthermore, although not perfect, the \ksmethod\ and \mlmethod\ produces more accurate estimations of the uncertainties of the estimated parameters than \linmix\ (\autoref{fig:err_acc}). For sample sizes between 300 and 1000, while the \ksmethod\ no longer demonstrates superior performance, the \mlmethod\ continues to outperform the other methods. When the sample size is limited (below 300), \mlmethod, \linmix, and \leopy\ exhibit comparable performance; however, the \mlmethod\ possesses the advantage of GPU acceleration.

While the experiments on mock and real data involve one-dimensional regression problems, the method itself is derived within the framework of multidimensional regression (\autoref{sec:intro_method}), where $\bx$ is always assumed to be a multidimensional variable. As NF can effectively model high-dimensional distributions, generalization to higher dimensions is straightforward. Most existing methods can be considered as special cases of the techniques we propose (\autoref{sec:work_compare}). We recommend using the \ksmethod\ as the first choice in general scenarios. However, when the sample size is limited ($N\lesssim 100$, see \autoref{fig:diff_N} and \autoref{sec:sample_size} for details), the \mlmethod\ is recommended as an alternative. We provide suggestions for handling the dependence of different data points and the existence of censored or missing data in practice in \autoref{sec:dicussion}.  Addressing these complex behaviors natively represents a key direction for future development. 

A GPU-compatible Python implementation of our method, nicknamed \raddest\ (Regression for Astronomical Data with realistic Distributions, ErrorS and non-lineariTy), will be released publicly upon acceptance of the paper\footnote{https://github.com/astro-jingtao/raddest}.

\section*{Acknowledgement}

We are grateful to the anonymous referee and AAS editors for their comments which have helped us to improve this paper. This work is supported by the National Natural Science Foundation of China (grant No. 12433003, 11821303, 11973030), and 
the National Key R\&D Program of China (grant No. 2022YFA1602902). TJ thanks Kai Wang, Yangyao Chen, Kangning Diao, Ce Sui, and Zechang Sun for helpful discussions. 

This paper makes use of the ALMA data, publically available at ADS/JAO.ALMA (\#2012.1.00650.S, \#2013.1.01161.S, \#2015.1.00925.S, \#2017.1.00392.S, \#2017.1.00886.L). 
ALMA is a partnership of ESO (representing its member states), NSF (USA) and NINS (Japan), together with NRC (Canada), MOST and ASIAA (Taiwan), and KASI (Republic of Korea), in cooperation with the Republic of Chile. The Joint ALMA Observatory is operated by ESO, AUI/NRAO and NAOJ. The National Radio Astronomy Observatory is a facility of the National Science Foundation operated under cooperative agreement by Associated Universities, Inc.

This work is supported by China Manned Space Program through its Space Application System. 

This work is based in part on observations made with the NASA/ESA/CSA James Webb Space Telescope. The data were obtained from the Mikulski Archive for Space Telescopes at the Space Telescope Science Institute, which is operated by the Association of Universities for Research in Astronomy, Inc., under NASA contract NAS 5-03127 for JWST. These observations are associated with program 2107.

The authors acknowledge the Tsinghua Astrophysics High-Performance Computing platform at Tsinghua University for providing computational and data storage resources that have contributed to the research results reported within this paper.

\appendix
\renewcommand\thefigure{\Alph{section}\arabic{figure}} 
\setcounter{figure}{0}

\section{The setting of $P(\log x)$ and the generation of uncertainty level} \label{app:d_n_setting}

\begin{figure*}[t!]
	\centering
	\includegraphics[width=0.8\textwidth]{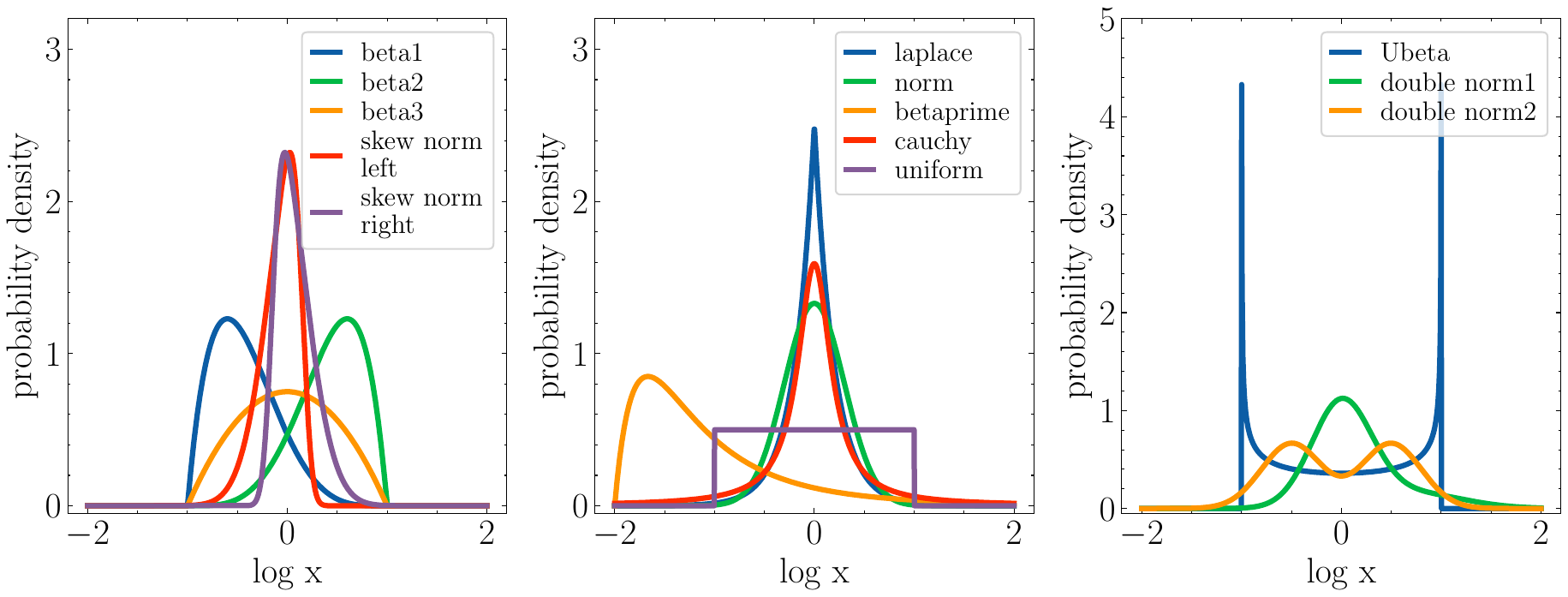}
 \caption{Left and center: 10 different predefined distributions of $P(\log x)$ used for all samples except Sample 3. Right: 3 different predefined distributions of $P(\log x)$ used for Sample 3. \label{fig:dist}}
\end{figure*}

The intrinsic distribution of $\log x$, denoted as $P(\log x)$, is randomly selected from ten predefined distributions shown in left and center panels of \autoref{fig:dist} for the general case. For datasets in sample 3, $P(\log x)$ is randomly chosen from three more complex distributions illustrated in right panel of \autoref{fig:dist}.

\begin{figure*}[ht!]
	\centering
	\includegraphics[width=\textwidth]{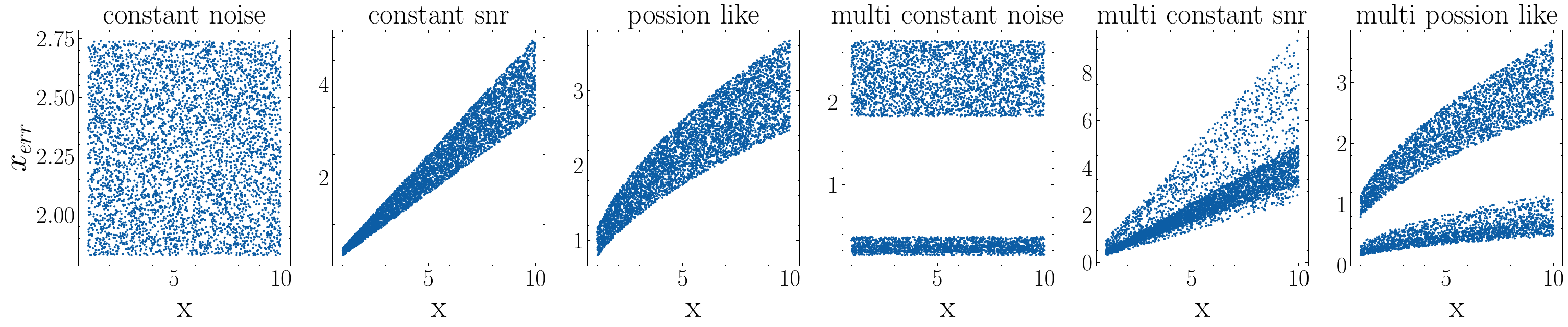}
	\caption{The examples of different correlations between $x$ and $\err{x}$.\label{fig:x_xerr}}
\end{figure*}

Noise is generated for both $x$ and $y$ using the same methodology. Therefore, we only describe the method employed to generate the noise for $x$ below. For each dataset, the typical S/N of $x$ ($x_{S/N}$) is randomly drawn from a uniform distribution between 1 and 10. The parameters $\err[,scale]{x}$, controlling the diversity of $\err{x}$ at a given $x$, are randomly selected from a uniform distribution between 0.1 and 0.9. The basic uncertainty level, $\err[,level]{x}$, for each data point is then generated from a uniform distribution between ${\rm MAX}(1/x_{S/N} - 0.5\ \err[,scale]{x}, 0)$ and $(1/x_{S/N} + 0.5\ \err[,scale]{x})$. 

Next, the correlation between $x$ and $\err{x}$ is randomly chosen from ``constant noise'', ``constant S/N'', and ``Poisson-like''. The uncertainty level, $\err{x}$, is calculated according to \autoref{eq:const_noise}, \autoref{eq:const_snr}, and \autoref{eq:possion_like} for ``constant noise'', ``constant S/N'', and ``Poisson-like'', respectively. 

\begin{equation} \label{eq:const_noise}
	\err{x} = \err[,level]{x} \times {\rm MEDIAN}(x)
\end{equation}

\begin{equation} \label{eq:const_snr}
	\err{x} = \err[,level]{x} \times x
\end{equation}

\begin{equation} \label{eq:possion_like}
	\err{x} = \err[,level]{x} \times \sqrt{x \times {\rm MEDIAN}(x)}
\end{equation}

The procedure outlined above is applied to datasets from all samples except sample 4, which is discussed separately below.

For datasets in sample 4, the basic uncertainty level, $\err[,level]{x}$, for each data point is generated from a uniform distribution between 1/6 and 1/2. The correlation is randomly selected from ``multi constant noise'', ``multi constant S/N'', and ``multi Poisson-like''. First, $\err{x}$ is calculated using \autoref{eq:const_noise}, \autoref{eq:const_snr}, and \autoref{eq:possion_like} for ``multi constant noise'', ``multi constant S/N'', and ``multi Poisson-like'', respectively. Subsequently, 10\% of $\err{x}$ values are randomly selected and scaled by a number randomly drawn from the interval [0.5, 2]. This random scaling process is repeated five times. 

Examples of different correlations between $x$ and $\err{x}$ are presented in \autoref{fig:x_xerr}. 

\section{The Saturated Log-linear Correlation} \label{app:saturate}

A log-linear correlation can be expressed as 
\begin{equation}
    \log y = k \log x + b
\end{equation}
where $k$ is the slope and $b$ is the intercept.

The saturated version of this correlation is defined as follows:

\begin{equation}
    \log y = 
    \begin{cases} 
        m_0 e^{n_0 \log x} + c_0 & \text{if } \log x < \log x_0 \\ 
        k \log x + b & \text{if } \log x_0 \leq \log x \leq \log x_1 \\ 
        m_1 e^{n_1 \log x} + c_1 & \text{if } \log x > \log x_1 
    \end{cases}
\end{equation}
where 
\begin{itemize}
    \item $\log x_0$ and $\log x_1$ are boundary points that define different growth patterns across varying intervals of the function. These two points are set as 5\% and 95\% quantiles of $\log x$, respectively.
    \item $c_0$ and $c_1$ are constants that represent the asymptotic values under the saturation conditions at both ends. They are set as 1\% and 99\% quantiles of $k \log x + b$, respectively.
    \item $n_0 = k / (k \cdot x_0 + b - c_0)$, $n_1 = k / (k \cdot x_1 + b - c_1)$
    \item $m_0 = k / (n_0 \cdot e^{n_0 \cdot x_0})$, $m_1 = k / (n_1 \cdot e^{n_1 \cdot x_1})$
\end{itemize}

\begin{figure*}[ht!]
	\centering
	\includegraphics[width=0.65\textwidth]{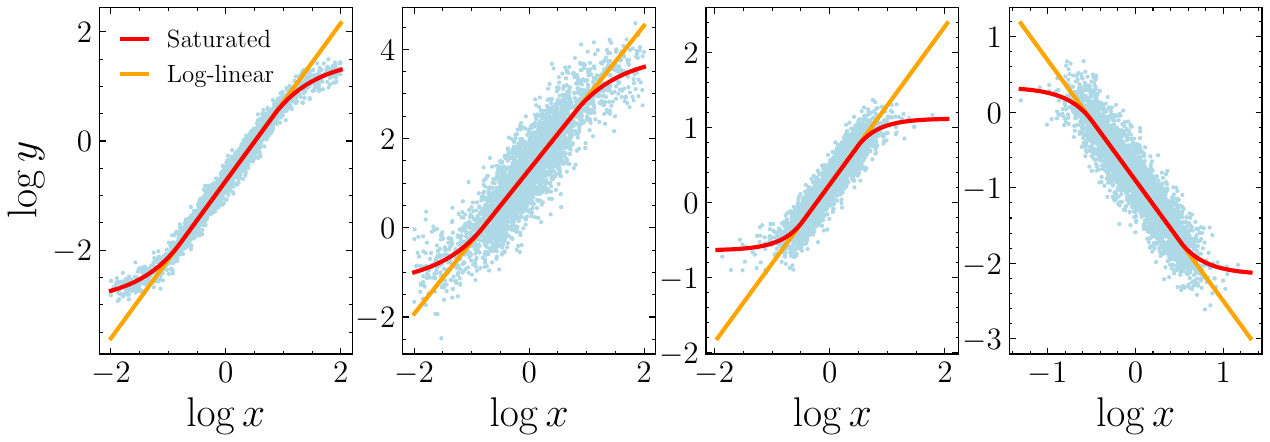}
	\caption{Comparisons between log-linear correlations (orange lines) and their saturated versions (red lines). The light blue dots represent the data points with saturated log-linear relationship and intrinsic scatter. \label{fig:example_saturation}}
\end{figure*}

\autoref{fig:example_saturation} illustrates a comparison between log-linear correlations and their saturated counterparts. 

\bibliography{sample631}{}
\bibliographystyle{aasjournal}

\end{document}